\documentclass[article,nofootinbib]{revtex4}

\usepackage{graphicx}
\usepackage{epsfig}
\usepackage{fancyhdr}
\usepackage{empheq}
\usepackage{mathbbol}
\usepackage{wasysym}
\usepackage{bbm}
\usepackage{bm}
\usepackage{color}
\usepackage{amssymb,amsmath}
\usepackage{epstopdf}
\usepackage{float}
\usepackage{color}
\usepackage{subfig}

\graphicspath{{./png/}}

\newcommand{\be}{\begin{equation}}
\newcommand{\ee}{\end{equation}}
\newcommand{\bea}{\begin{eqnarray}}
\newcommand{\eea}{\end{eqnarray}}

\begin{document}

\title{SELECTED TOPICS IN MAJORANA NEUTRINO PHYSICS}
\author{Luciano MAIANI }

\address{ Dipartimento di Fisica, Universit\`a di Roma "La Sapienza", and INFN, Sezione di Roma, \\Piazzale A. Moro 5, Roma, I-00185, Italy.}


\begin{abstract}
Starting from the original Majorana's article of 1937, the see-saw mechanism is illustrated, first for one and later for three neutrino generations, and neutrinoless double beta decay is considered. Neutrino mixing and oscillations in three flavors are described. The Yukawa couplings to the Higgs field of quarks and leptons are considered, their transformation properties under the corresponding flavor groups are spelled out and the principle of Minimal Flavor Violation is illustrated, in connection with possible new physics beyond the Standard Theory. The idea that the Yukawa couplings may be the vacuum expectation value of some new fields is introduced and natural extrema of potentials which are invariant under quark and lepton flavor groups are characterized. A recent  result indicating large mixing of almost degenerate neutrinos is derived from the heavy lepton invariance under flavor ${\cal O}(3)$.

\noindent
--------------------
 \newline 
 

 \end{abstract}
 

\maketitle

 {\bf
SUMMARY
\begin{enumerate}
\item  The Majorana Neutrino
\item See-Saw Mechanism and Double Beta Decay without Neutrinos
\item Quark Masses and Mixing
\item FCNC Processes: Standard Theory and Beyond
\item Neutrino Mixing and Oscillations with Three Flavors
\item See-saw Neutrinos in Three Generations
\item Yukawa Couplings as Fields
\item Outlook
\end{enumerate}
}

\section*{ Foreword}

The subjects of these notes originate from three lectures given at the Physics Department of University Federico II, Naples, in spring 2014. The title, Ettore Majorana Lectures, itself invited to speak about neutrinos, indeed a very fascinating subject in these times.

The present Standard Theory is based in great part on the concept of Quark-Lepton Universality of the dominant gauge interactions. However, theoretical and experimental work accumulated over the last decades strongly indicates that we have to depart from the concept of a strict analogy between quarks and leptons, if we want to understand the pattern and the origin of their masses and mixing angles, what is called, in brief, Flavor Physics.  

Neutrinos have masses ways below the other fermion masses and the surprise of the last years has been that, unlike quarks, their mixing angles are generally large. A Majorana neutrino has been invoked to explain the first aspect, the so-called {\it see-saw mechanism}, and a large experimental effort is being put in the search for positive evidence of neutrinoless double beta decay ($\beta\beta0\nu$ decay).

To explain the large neutrino mixing angles, new concepts such as the invariance under discrete symmetries have been invoked. I shall report on a recent work where the large mixing angles arise from another characteristic of Majorana neutrinos, namely to have as  flavor symmetry orthogonal rather than unitary groups. The approach we propose brings another surprise, namely it leads to almost degenerate neutrinos. If true, this would be a dramatic departure from the quark-lepton analogy and, most important, it would put the neutrinoless double beta decay at a level not far from present experimental limits.


In what follows, starting in Sect.~\ref{majonu} from the original Majorana paper~\cite{majorana1,majorana2}, I shall illustrate the see-saw mechanism~\cite{Minkowski:1977sc,GellMann:1980vs,yanagida,glashowseesaw,mohapatra} and the implied $\beta\beta0\nu$ decay, Sect.~\ref{ssmech}, in the case of one neutrino flavor. 

Quark masses and mixing are introduced in Sect.~\ref{qm&m}. A discussion of Flavor Changing Neutral Current processes is given in Sect.~\ref{fcnc}, with a presentation of the quark flavor group and the related Minimal Flavor Violation principle~\cite{mfv}. 

Sect.~\ref{num&m} presents a brief review of neutrino oscillations, including the recent determination of the last real mixing angle, $\theta_{13}$, at the Daya Bay reactor~\cite{dbchina} and the direct evidence for $\nu_\mu \to \nu_\tau$ oscillations obtained by the Opera Collaboration~\cite{Agafonova:2010dc}. The lepton flavor group with three generations and the corresponding see-saw mechanism  are illustrated in Sect.~\ref{seesawnu}. 

The idea that Yukawa couplings may be the vacuum expectation value of some new fields is finally introduced in Sect.~\ref{dynyuk}, where we characterize the {\it natural extrema} of potentials which are invariant under the quark and the lepton flavor groups. This will make it possible to derive the announced result of large mixing of almost degenerate neutrinos~\cite{Alonso:2013nca}, to be compared with the hierarchical masses and small mixing angles found for quarks.

Open problems are recalled in Sect.~\ref{outlook}.

\section{ The Majorana Neutrino}
\label{majonu}

In 1937 Majorana wrote a paper on the theory of electrons and positrons~\cite{majorana1}, whose starting point was a reconsideration of the sea of negative energy states postulated by Dirac. 
At that time, the Dirac's sea was becoming a rather embarrassing object. The discovery of new particles implied a different sea for each particle. 
Moreover, there was no place for bosons because the key idea of the Dirac theory rested in the fact that the sea was filled by particles obeying the Pauli exclusion principle. 

Majorana set up to eliminate this sort of ``ether''. 

\subsection{\bf A symmetric theory of electrons and positrons}

In Dirac's theory, the interpretation of the negative energy states leads to a symmetric description of the electrons and the positrons. But this symmetry is not 
evident at all at the beginning. 
There exist in fact examples in the physics of solids where there are bands almost completely filled and where electrons and holes do not have the same mass. 

Interestingly, Dirac himself had originally speculated that the 
mass of the hole could be different from the mass of the electron and that, perhaps, the hole could correspond to the proton.
It was only after H. Weyl demonstrated formally the symmetry under charge conjugation of the basic electrodynamics that it was 
understood that the positron had to have the same mass of the electron. 

As Majorana noticed
\footnote{Quotations in english are taken from my translation~\cite{majorana2} of Majorana's paper. The corresponding quotation from the original paper \cite{majorana1} is  :\textit{``Tuttavia gli artifici suggeriti per dare alla teoria una forma simmetrica che si accordi con il suo contenuto, non sono 
del tutto soddisfacenti; sia perch\'e si parte sempre da una impostazione asimmetrica, sia perch\'e la simmetrizzazione viene in seguito ottenuta mediante tali 
procedimenti (come la cancellazione di costanti infinite) che possibilemte dovrebbero evitarsi. Perci\'o abbiamo tentato una nuova via che conduce pi\'u 
direttamente alla meta.''}}:
\\
\indent
\textit{``The prescriptions needed to cast the theory into a symmetric form, in conformity with its content, are however not entirely satisfactory, either because one 
always starts from an asymmetric form and because symmetric results are obtained only after one applies appropriate procedures, such as the cancellation of 
divergent constants, that one should possibly avoid. For these reasons, we have attempted a new approach, which leads more directly to the desired result.''
}

The procedure he suggested is essentially what we know now as Quantum Field Theory which provides a unique vacuum (no sea of negative energy states) 
and particles which are excitations of this vacuum. 
Majorana then observes\footnote{\textit{``Per quanto riguarda gli elettroni e i positroni, da essa si pu\'o veramente attendere soltanto un progresso formale; ma ci sembra importante, per le possibili astensioni analogiche, che venga a cadere la nozione stessa di stato di energia negativa. Vedremo infatti che \'e perfettamente possibile costruire, nella maniera pi\'u naturale, una teoria delle particelle neutre elementari senza stati negativi.'}'}:
\\
\indent
\textit{``In the case of electrons and positrons, we may anticipate only a formal progress; but we consider it important, for possible extensions by analogy, that the very notion of negative energy states can be avoided. We shall see, in fact, that it is perfectly, and most naturally, possible to formulate a theory of elementary neutral particles which do not have negative (energy) states.''}

The surprise 
was that a description of a spin 1/2 particle was possible, which involves only 2 degrees of freedom (spin up and 
spin down) and not 4 as in Dirac's theory. Such a particle is neutral, in the sense that  it coincides with its antiparticle, and it corresponds to the {\it Majorana neutrino}. 

To construct his theory, Majorana used a representation where the Dirac matrices are all imaginary (since known as the Majorana representation, MR). In this representation the Dirac equation
\begin{equation}
\left( i \gamma^{\mu} \frac{\partial }{\partial x^{\mu}}+m \right) \psi(x)=0,
\end{equation}
has real coefficients. Therefore, setting 
\begin{equation}
\psi(x)= U(x)+iV(x), \label{Majorana U V}
\end{equation}
 $U$ and $V$ never mix~\footnote{\textit{``Ma \'e notevole che la parte di tale formalismo che si riferisce alle 
$U$ (o alle $V$) possa \textit{da sola} essere considerata come descrizione teorica, in armonia con i metodi generali della meccanica quantistica, di un qualche 
sistema materiale.''} }:
\\
\indent
\textit{``It is remarkable, however, that the part of the formalism which refers to $U$ (or $V$) can be considered, in itself, as the theoretical description of some material system, in conformity with the general methods of quantum mechanics.''}

In a normal representation, if we start with a real wave function, the time evolution makes it complex because the equation has complex coefficients, but
in the Majorana representation if we start from $U$, with  $V$ equal zero, $\psi$ remains real and it gives an acceptable description of some material system.

Majorana promptly recognized that one needs to introduce both $U$ and $V$ to describe the electron which is a particle that admits a conserved charge. 
However, the simplicity of the scheme leads him to speculate that his theory can be applied to electrically neutral particles\footnote{ 
\textit{``Il fatto che tale formalismo ridotto non si adatti alla descrizione degli elettroni positivi e negativi, pu\'o bene essere dovuto alla presenza della carica elettrica e non impedisce l'affermazione che allo stato attuale delle nostre conoscenze le (12) e (13) costituiscono la pi\'u semplice rappresentazione teorica di un sistema  di particelle neutre. Il vantaggio di questo procedimento rispetto all'interpretazione elementare delle equazioni di Dirac \'e (come vedremo meglio fra poco) che non vi \'e pi\'u nessuna ragione di presumere l'esistenza di antineutroni o antineutrini. Questi ultimi vengono in realt\'a utilizzati nella teoria dell'emissione $\beta$ positiva, ma tale teoria pu\'o essere, ovviamente, modificata in modo che l'emissione $\beta$, sia negativa che positiva, venga sempre accompagnata dall'emissione di un neutrino.''}}:
\\
\indent
\textit{``The fact that the reduced formalism cannot be applied to the description of positive and negative electrons may well be attributed to the presence of the 
electric charge, and it does not invalidate the statement that, at the present level of knowledge, eqs. (12) and (13) constitute the simplest theoretical representation of neutral particles.}[note: numbers refer to the equations in the original paper which characterize the Majorana fermion]  {\it The advantage, with respect to the elementary interpretation of the Dirac equation, is that there is now no need to assume the existence of 
antineutrons or antineutrinos (as we shall see shortly). The latter particles are indeed introduced in the theory of positive $\beta$-ray emission; the theory, 
however, can be obviously modified so that the $\beta$-emission, both positive and negative, is always accompanied by the emission of a neutrino.''}

Majorana refers here to the theory of positive $\beta$-rays formulated two years before, in Rome, by Giancarlo Wick~\cite{wick}.

The Majorana scheme represents in fact the simplest theoretical description of a neutral, spin $1/2$ particle.

\subsection{\bf Pontecorvo, Fermi and  Don Quixote }
In 1934, Hans Bethe and Rudolf Peierls computed the probability for a neutrino to be detected by its interaction with matter in the inverse process of  the beta decay~\cite{bethe1}:
\begin{equation}
\bar{\nu}_e+p \rightarrow e^{+}+n. \label{inverse beta raction}
\end{equation}

Under the condition $E_{\nu}\ll M$, with $E_{\nu}$ the neutrino energy and $M$ the nucleon mass, they found that the cross-section for the interaction of a neutrino with a nucleus is approximatively given by $G^{2}E_{\nu}^{2}$, where $G$ is the Fermi constant. 
This implies that the mean free path of $1$~MeV neutrino in iron ($\rho_{iron}\approx 8$~gr/cm$^{3}$) is approximately
\begin{equation}
L\approx 6~ {\rm light~years} \cdot \frac{1}{\left[ E_{\nu}(MeV) \right]^2},
\end{equation}
or in other words that the probability of interaction in $l=1$~m of iron is:
\begin{equation}
P\approx 2\cdot 10^{-17} \left[ E_{\nu}(MeV) \right]^2.
\end{equation}

This result discouraged for many years all attempts to observe the neutrinos, until Pontecorvo realized in 1947 that, although the probability of interaction of the neutrino is astronomically small, a nuclear reactor produces an equally astronomical quantity of neutrinos. A nuclear reactor gives order of $10^{20}-10^{23}$ neutrinos per second, so that in an iron of length $l=1$~m we could have as many as $N \approx 10^3$~events per second. While in Canada, Pontecorvo devised  {\it radiochemical methods} to reveal neutrinos from a nuclear reactor or from the Sun.

Immediately after, he made a trip to Europe and talked about his method  to Pauli, who was interested. Then he talked to Fermi, who, on the contrary, did not show much interest, probably thinking that it would take decades to develop the method completely.  

Everybody has his/her own heroes. Emilio SegreÕ, in this connection, notes: {\it Don Quixote was not a Fermi's hero}.  But Pontecorvo's paper had the virtue to reopen the issue of neutrino's experimental observation.

\paragraph{{\bf On Pontecorvo's method} } Pontecorvo proposed what is now called the Chlorine-Argon method. A tank of Chlorine atoms is exposed to neutrinos, which induce the reaction: $\nu + Cl^{37}\to Ar^{37}+e^-$. The produced Argon nucleus decays back to a Chlorine nucleus by beta decay:  $Ar^{37}\to Cl^{37}+ \nu + e^+$, with a lifetime of $34.3$~days. There is therefore time to extract the Argon from the tank by bubbling air in it, to collect it in a separate vessel and to measure the number  of Argon atoms produced, say, in one day, by measuring its radioactivity. Knowing the cross section, we can measure in this way the neutrino flux. In the Fermi theory the particles produced by the nuclear reactor are in fact {\it antineutrinos}, arising from neutron's beta decay: $n \to p+{\bar \nu} + e^-$. However, in these times  the concept of lepton number conservation was no so well established. In addition, following Majorana,  one could think that the method would work if neutrinos and antineutrinos are the same particle.  Pontecorvo considered the alternative transition  $Cl^{35} \to S^{35}$, which in Fermi's theory would be produced by antineutrinos according to ${\bar \nu} + Cl^{35} \to S^{35}+ e^{+}$. The $Cl^{37}\to Ar^{37}$ transition is useful for {\it nuclear fusion} reactions, where protons are fused into He nuclei, with emission of $e^+ \nu$ pairs. As such, it has been employed  by R. Davis to detect  {\it solar neutrinos}, see Sect.~\ref{num&m}.  Much later, the $Ga^{71} \to Ge^{71}$ transition induced by solar neutrinos has also been considered. No use has been proposed, untill now, of the $Cl^{35} \to S^{35}$ transition. See ~\cite{fidecaro} for a very accurate discussion of Pontecorvo's radiochemical methods.

 \subsection{{\bf The observation of neutrinos}} 
\begin{figure}[htbp]
       \centering
        {
        \includegraphics[scale=0.33]{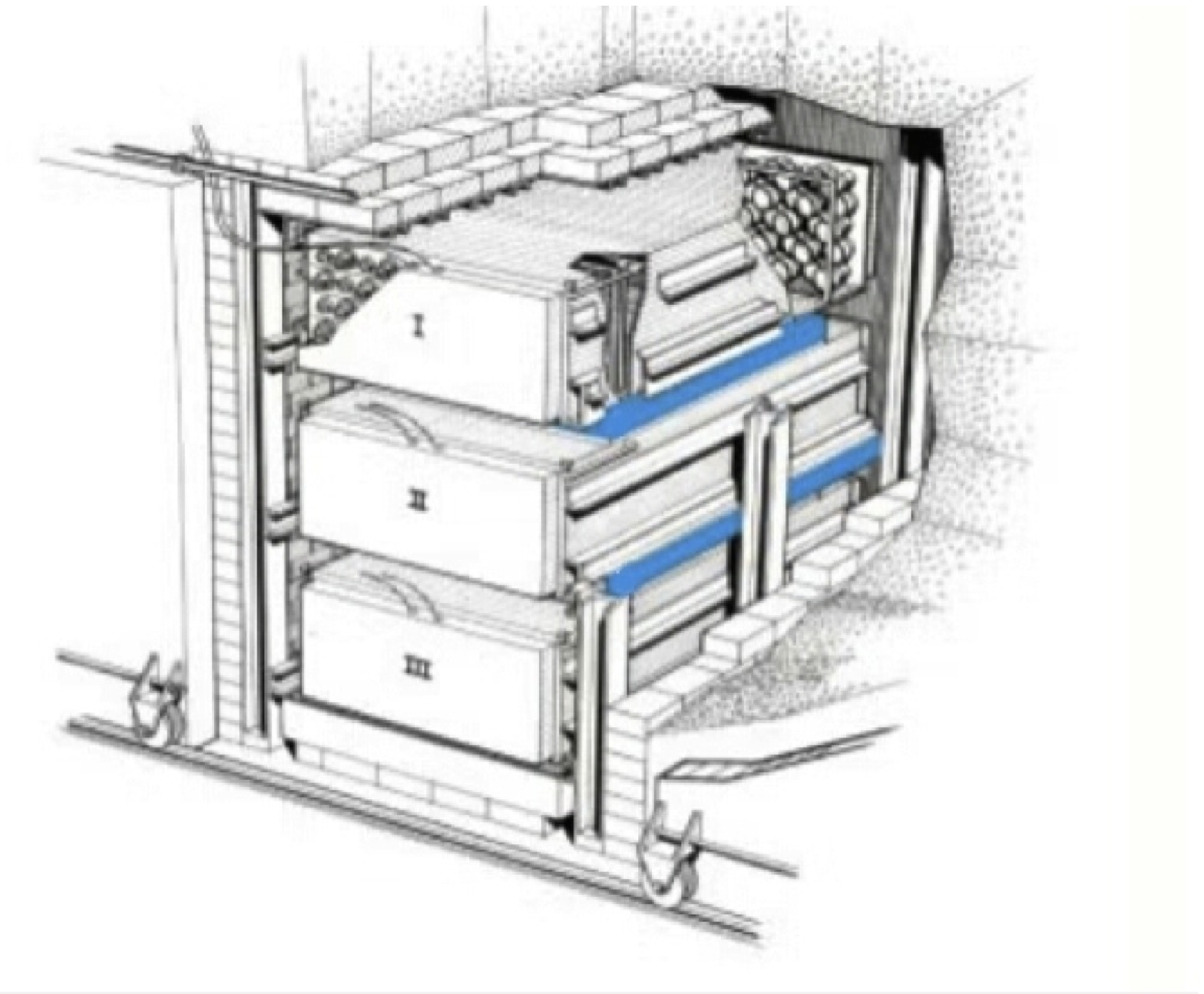}}
       {
        \includegraphics[scale=0.33]{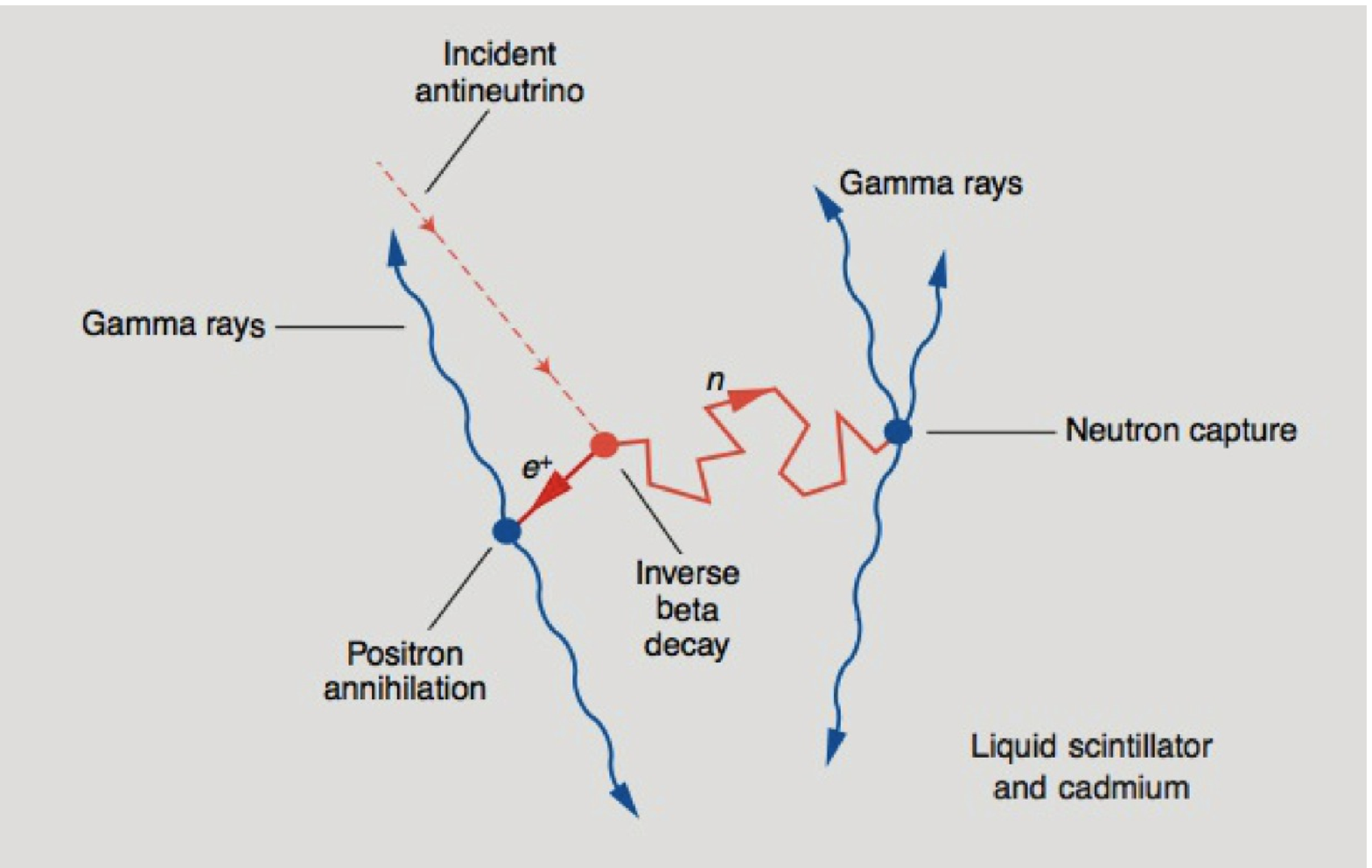}}
        \caption{Tanks I, II and III of the Reines and Cowan detector were filled with liquid scintillator and instrumented with 5'' PMTs.
        Target tanks (in blue) were filled with water+cadmium chloride. Inverse $\beta$ decay would produce two signals in neighbouring tanks (I, II or
       II, III): a prompt signal from $e^+$ annihilation producing two $0.51$~MeV $\gamma$s and a delayed signal from $n$ capture on 
          cadmium producing $9$~MeV $\gamma$s. Figures from {\it Los Alamos Science}~\cite{losalamos}. }
        \label{fig:detectorR&C}
\end{figure}

In 1953 F. Reines and C. Cowan proposed a detector for antineutrinos produced by the Savannah River reactor in the inverse beta reaction (\ref{fig:detectorR&C}).
The experiment did not give a definite  response, as the background represented by cosmic rays was still very large and produced events similar to the  ones given by (\ref{inverse beta raction}). 

In 1956 Reines and Cowan set up a completely revised detector, see Fig.\ref{fig:detectorR&C}.

The detector was composed by three tanks filled scintillator liquid. The scintillator tanks sandwitched water tanks containing cadmium chloride that is a highly effective neutron absorber with the emission of gamma rays. 

A neutrino interacting with protons in one water tank creates a neutron and a positron. The positron gives rise to a pair of gamma rays when it annihilates with an electron of the water and the neutron gives delayed gamma rays when captured by cadmium. Gamma rays are detected by  the photomultipliers observing the liquid scintillators in  two tanks at most, i.e. those sandwiching the water tank, but not in three, as most of the spurious cosmic ray signals would  do. 

With this clever discrimination of background signals due to cosmic rays, Reines and Cowan could indeed announce in 1956 the observation of unequivocal signals of the interaction of antineutrinos produced by the reactor~\cite{reines}, with a frequency compatible with the prediction of Bethe and Peierls.

\subsection{\bf Surviving the data} In the Fermi theory,  neutrinos are Dirac particles associated to a conserved charge, i.e. lepton number, and the basic transitions are:
\begin{eqnarray}
&&n \rightarrow p + e^{-} + \bar{\nu},~{\rm or}~\nu+n\to p+e^-\\
&& p \rightarrow n + e^{+} + \nu,~{\rm or}~\bar \nu +p \to n+e^+.
\end{eqnarray}
(the $\beta^+$ decays occur in isotopes where the mass difference beteween neutron and proton is compensated by the reducton of the electrostatic repulsion due to the disappearence of the proton). 

Lepton number conservation is observed in the chain of production and subsequent reaction, in the sense that the particle produced in $\beta$-decay in association with the electron (the antineutrino, in Fermi's theory)  produces a positron in its subsequent interaction and never an electron. Thus, we would conclude that neutrinos cannot be Majorana particles, because the neutrino which is emitted together with the electron is different from the one emitted with the positron, i.e. $\nu \neq \bar{\nu}$. However, this conclusion does not hold for very light neutrinos in the presence of maximal parity violation: field theory and Nature are smarter. 

If we combine Majorana theory with V-A interaction, $\beta^{-}$ and $\beta^{+}$ emission are respectively described by the current (we use the MR, where $\gamma_0$ and  $\gamma_5$ are both imaginary and antisymmetric) 
\begin{equation}\label{betadecaycurrent}
{\it J}^\mu={\bar \psi_e} \gamma_\mu \frac{1}{2}(1-\gamma_5) U={\bar \psi_{e_{L}}} \gamma_\mu U_L,
\end{equation}
and its hermitian conjugate:
\begin{equation}
({\it J}^\mu)^\dagger=U^T\gamma^0\gamma_\mu \frac{1}{2}(1-\gamma_5) \psi_e=(\frac{1}{2}(1+\gamma_5) U)^T\gamma^0 \gamma_\mu \psi_{e_{L}}=(U_R)^T\gamma^0\gamma_\mu \psi_{e_{L}},
\end{equation}
$U$ being the Majorana field introduced in (\ref{Majorana U V}).  Thus, the neutrino produced in $\beta^-$ decay has positive chirality, while the one of the $\beta^{+}$ decay has negative chirality. 

Now, for processes involving very light particles such as neutrinos and Vector or Axial vector interaction, chirality is almost equal to helicity which, in turn, is almost exactly conserved. Therefore the neutral particles produced in these two processes, despite being two components of the same Majorana field, ARE different.

We can say that the role played by the lepton number in Fermi's theory is taken over by helicity. The latter is exactly conserved for massless particles, while for massive particles, the violation of lepton number arises only to order $(m_{\nu}/E_{\nu})^2$, which is a neglegible effect as $E_{\nu} \sim {\rm MeV}$ in $\beta$ decays and $m_{\nu} \ll E_{\nu}$. 

As was realized in the late fifties, a massless Majorana neutrino with V-A interaction is mathematically equivalent to the two component Weyl neutrino. 

\subsection{\bf Weyl, Majorana and Dirac neutrinos} 
\begin{figure}[htbp]
       \centering
                \includegraphics [scale=0.5]{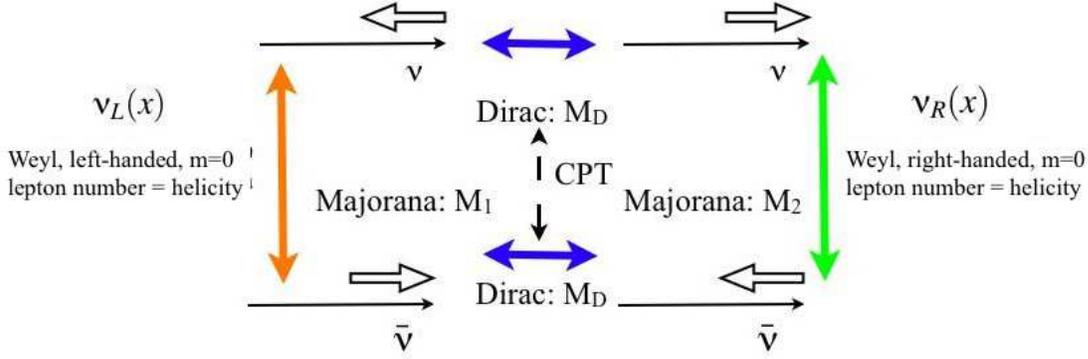}
                              \caption{Weyl, Majorana and Dirac neutrinos, see text. Figure from Ref.~\cite{benhar1}} 
        \label{fig:WMD}
\end{figure}

A comparison of the Weyl, Majorana and Dirac theories is made in Fig.~\ref{fig:WMD} (see Ref.~\cite{benhar1}).

In Dirac's theory the neutrino is a four dimensional Dirac spinor field, whose positive  frequency parts annihilate neutrino states with the helicity $\pm 1/2$, shown in the upper part of Fig.~\ref{fig:WMD}, while the negative frequency part create antineutrino states with helicity $\pm 1/2$, lower part of Fig.~\ref{fig:WMD}. In the zero mass limit, the Dirac field decomposes into a pair of two-dimensional Weyl fields, indicated by the vertical arrows. More precisely:
\begin{itemize}
\item 
a Weyl left-handed spinor field, $\nu_L$ (Fig.~\ref{fig:WMD} left):  the positive frequency part annihilates a neutrino state with negative helicity and the negative frequency part creates an antineutrino state with positive helicity.

\item
a Weyl right-handed spinor field, $\nu_R$ (Fig.~\ref{fig:WMD} right):  the positive frequency part annihilates a neutrino state with positive helicity  and the negative frequency part creates an antineutrino state with negative helicity.
\end{itemize}

In the massless case, one could make a Majorana spinor using the same components as a left- or right-handed Weyl 
spinor: what was the lepton number in Weyl would be the helicity in Majorana.

A Dirac mass connects neutrino states horizontally, thereby respecting the lepton number, while a Majorana mass connects the two Weyl states on the left, or the two states on the right, thereby violating the lepton number. 

We may translate in formulae the content of Fig.~\ref{fig:WMD} as follows.

Starting from the two Weyl fields $\nu_L$ and $\nu_R$, we obtain  {\it {\bf two real Majorana fields}} according to (Majorana representation used throughout):
\begin{equation}
\psi_1 = \nu_{L} + \left( \nu_L \right)^{\star},
\label{majo1}
\end{equation}
\begin{equation}
\psi_2 = \nu_R + \left( \nu_R \right)^{\star},
\label{majo2}
\end{equation}
or {\it {\bf one Dirac spinor}} according to:
\begin{equation}
\psi_D = \nu_L + \nu_R.
\label{dirac}
\end{equation}
Masses can be given to $\psi_{1,2}$ with the mass lagrangians:
\begin{equation}
{\cal L}_{mass,1} = \dfrac{1}{2} M_1 \psi_1^T \gamma^0 \psi_1 + {\rm h. c.},\label{Mmass1}
\end{equation}
\begin{equation}
{\cal L}_{mass,2} = \dfrac{1}{2} M_2 \psi_2^T \gamma^0 \psi_2 + {\rm h. c.},
\label{Mmass2}
\end{equation}
$M_1$ and $M_2$ are obviously called Majorana masses. 

In the same notation, a Dirac mass takes a non diagonal form in $\psi_1$ and $\psi_2$: 
\begin{equation}
{\cal L}_{mass,D} = M_D{\bar \psi_D}\psi_D=M_D[{\bar \nu_L}\nu_R~+~{\rm h.c.}]=\frac{1}{2}M_D[\psi_1^T\gamma^0\psi_2 + \psi_2^T\gamma^0 \psi_1]\label{Dmass}
\end{equation}
where $M_D $ is the Dirac mass. 
\paragraph{\bf Note} {\small The notation used in eqs.~(\ref{Mmass1}) and (\ref{Mmass2}) is such that $\psi_1$ denotes a column vector and $\psi_1^T$ a row vector, so that the product is executed with the row-times-colum rule; for simplicity of notation, we shall often omit the transpose symbol, understanding that all vector and matrix products are executed with this rule}.
\vskip0.5cm

The matrices which represent the generators of the Lorentz transformations in Dirac's theory are:
\begin{equation}\nonumber
\sigma_{\mu\nu}=\frac{i}{2}\left[\gamma^\mu,\gamma^\nu\right]
\end{equation}
which are imaginary matrices in the MR, such that:
\begin{equation}\nonumber
\sigma_{\mu\nu}^T\gamma^0= - \gamma^0 \sigma_{\mu\nu}
\end{equation}

This is as it should be, in order for the matrices representing finite Lorentz transformations to be real, pseudo-orthogonal  matrices:
\begin{equation}
S(\Lambda)=e^{i \alpha_{\mu\nu}\sigma_{\mu\nu}},~S^T\gamma^0=\gamma^0S^{-1}
\label{pseudo}
\end{equation}
and the components $U$ 
in (\ref{Majorana U V}) to be transformed among each other ($\alpha_{\mu\nu}$ are the real parameters characterizing the Lorentz transformation $\Lambda$).

As a consequence of (\ref{pseudo}), mass lagrangians in the Majorana form are Lorentz scalars, e.g.:
\begin{eqnarray}\nonumber
&&\psi_1(x)\to S(\Lambda)\psi_1(\Lambda^{-1}x); \\
\nonumber && {\cal L}_{mass,1} (x)\to [(S\psi_1)^T\gamma^0S\psi_1](\Lambda x^{-1})=[\psi_1^T S^T\gamma^0 S\psi_1](\Lambda x^{-1})=\\
\nonumber &&=[\psi_1^T \gamma^0 S^{-1} S\psi_1](\Lambda x^{-1})={\cal L}_{mass,1} (\Lambda x^{-1})
\end{eqnarray}
and the same for ${\cal L}_{mass,2}$ and ${\cal L}_{mass,D}$.

\newpage

\section{{ See-Saw Mechanism and Double Beta Decay without Neutrinos}}
\label{ssmech}

\subsection{\bf Elementary fermions}

\begin{figure}[htbp]
        \centering
       \includegraphics[scale=0.5]{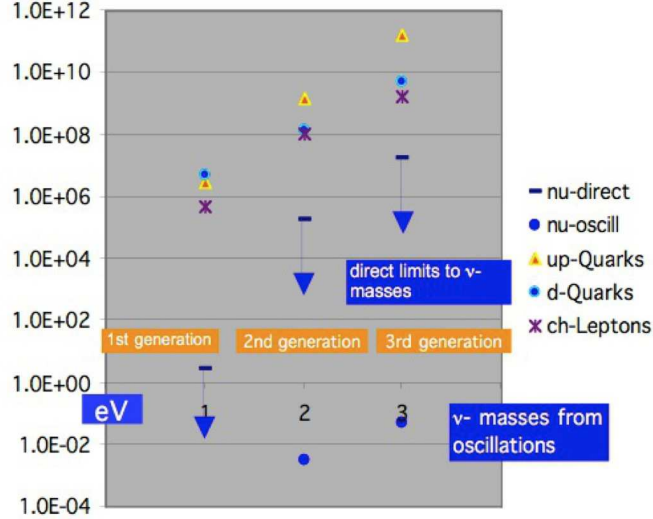}
  	   \caption{Masses of the elementary fermions (in eV). The upper ends of the vertical arrows indicate the experimental bounds to neutrino masses, obtained from beta decay spectra. Neutrino oscillations give values for the differences of the eigenvalues of the neutrino mass matrix, $m_2^2-m_1^2$ and $m_3^2-m_2^2$. Under the assumption of hierarchical neutrino masses: $m_3>>m_2>>m_1$, the masses of  $m_2$ and $m_3$ are indicated by the blue dots and $m_1<<m_2$. For hierarchical neutrino masses see, however, Sect.~\ref{dynyuk}.
}
  	   \label{fig:spectrum}
\end{figure}

A look at the mass spectrum of elementary fermions, quarks and leptons, shows that neutrinos are indeed very peculiar, see Fig.~\ref{fig:spectrum}.  

First generation leptons are the electron and its neutrino, with direct limits on the mass of the neutrino around $1$~eV.
Slightly above $1$~ MeV, the up and down quark (down quark is heavier than up quark and this explains why
the neutron is heavier than the proton). The second generation is formed by the $\mu$ lepton (mass around $100$~MeV) and its neutrino (with
mass direct  limits larger than the electron neutrino but much smaller that the muon mass) and by the strange quark, almost degenerate with the muon, and the 
charm quark. The third generation is made by the $\tau$ lepton, its neutrino, the bottom quark around $5$~GeV, and the top quark around $170$~GeV.

Non vanishing, albeit very small, masses are indicated by the phenomenon of neutrino oscillations (to be discussed later). 
This phenomenon gives the difference of the mass square of 2nd generation and 1st generation neutrinos and 3rd generation minus 2nd 
generation neutrinos. Assuming a hierarchical model ($m_{\nu_1}<< m_{\nu_2}<<m_{\nu_3}$) the observed mass-squared differences reproduce the masses of 2nd and 3rd generations neutrinos indicated in Fig.~\ref{fig:spectrum}.

Between a third generation neutrino of mass of $10^{-2}$~ eV and the top quark mass of $10^{11}$~eV there are about $13$ orders of magnitude. It is very difficult to 
imagine that these masses have the same origin. It is more reasonable to think that there is a common source for the masses of the quarks and
charged leptons (the coupling to the Higgs boson) and a different source for the neutrinos. This idea brings us again to Majorana theory.

\subsection{\bf Majorana neutrinos come back: the see-saw mechanism}
\label{mechanismss}

In the $60$'s, no attention was paid to the issue of Majorana neutrino because everybody believed neutrinos to be massless.

The Standard Model changed the attitude in regard to neutrino masses. It was realized that:
\begin{itemize}
\item chiral symmetry is broken so there is no reason a priori to expect massless neutrinos,
\item Dirac neutrino mass requires a right handed neutrino which does not interact (it would be a {\it sterile} neutrino).
\end{itemize}

The charged lepton masses are due to the coupling of a left handed to a right handed lepton via the Higgs field. But if right-handed neutrinos exist, coupled to the left-handed ones via the Higgs field,  why neutrinos are so much lighter than  their charged lepton counterparts?

Majorana mass and weak isospin selection rules make it possible to find a natural explanation to the smallness of neutrino mass which is called the {\it{\bf see-saw 
mechanism}}~\cite{Minkowski:1977sc,GellMann:1980vs,yanagida,glashowseesaw,mohapatra}. 

To mantain full generality, we describe left and right handed neutrinos of one generation with the two Majorana fields $\psi_{1,2}$ of eqs.~(\ref{majo1}) and (\ref{majo2}) and write the mass lagrangian as:
\begin{equation}
{\cal L}_{mass}={\cal L}_{mass,1}+{\cal L}_{mass,2}+{\cal L}_{mass,D}
\end{equation}

We find:
\begin{eqnarray}
\nonumber &&{\cal L}_{mass,1}=\frac{1}{2}M_1 \psi_1^T\gamma^0\psi_1=\frac{1}{2}M_1[\nu_L^T\gamma^0\nu_L\;+\;h.c.]\\
\nonumber &&{\cal L}_{mass,D}=M_D \psi_1^T\gamma^0\psi_2=M_D[\nu_R^\dagger \gamma^0\nu_L\;+\;h.c.]\\
\nonumber &&{\cal L}_{mass,2}=\frac{1}{2}M_2 \psi_2^T\gamma^0\psi_2=\frac{1}{2}M_2[\nu_R^T\gamma^0\nu_R\;+\;h.c.]
\end{eqnarray}

We know that $\nu_{L}$ has weak isospin $I_3=+1/2$ so the term in the first line has weak isospin $I_3=\pm 1$ and it cannot be produced by a coupling with a $I=1/2$ Higgs doublet: we expect  $M_1= 0$.

The term  in the second line has weak isospin $I=\pm 1/2$ , so the mass $M_D$ can be produced by a coupling to the Higgs doublet, entirely similar to the charged lepton coupling: we expect  $M_D$ $\approx$ normal lepton and quark masses.

Finally, the term in the third line has  weak isospin $I=0$ and vanishing weak hypercharge; the mass $M_2$ can be anything since it does not break the gauge symmetry of the Standard Theory. Most naturally we expect $M_2\approx M_{GUT} \approx 10^{14-15}$ GeV.

Combining these considerations, we get the Majorana mass matrix:

\begin{equation}\label{Mmatrix}
\begin{pmatrix} 0 & M_D \\ M_D & M_{GUT}\end{pmatrix}
\end{equation}

For $M_D<<M_{GUT}$, this matrix has a small eigenvalue:
\begin{equation}\label{seesaw}
m_{\nu_L}\approx \frac{M_D^2}{M_{GUT}}
\end{equation}
corresponding to an almost pure $\psi_1$, and a large eigenvalue equal to $M_{GUT}$ corresponding to an almost pure $\psi_2$. 
In conclusion:
\begin{itemize}
\item $\nu_R$ exists with a mass $=M_{GUT}$,
\item $\nu_L$ acquires a Majorana mass $m_\nu=\frac{M_D^2}{M_{GUT}}$.
\end{itemize}

If we take $m_{top}$ as a natural value for the mass $M_D$ for the 3rd generation neutrino and $M_{GUT}= 10^{15}$ GeV, we get $m_{\nu}= 3\cdot 10^{-2}$ eV, remarkably close to the estimate reported in Fig.~\ref{fig:spectrum}, based on the SuperKamiokande results from the oscillation of atmospheric neutrinos.

Of course the eigenvectors of (\ref{Mmatrix}) are Majorana neutrinos. The heaviest one is essentially a right handed neutrino and the lowest one is essentially
the left handed neutrino \textit{\'a la} Majorana with a small mass.

The see-saw result in (\ref{seesaw}) can be also obtained by an independent line of reasoning due to S. Weinberg~\cite{Weinberg:1979sa}.

As noted, the lagrangian ${\cal L}_{mass,1}$ has weak isospin $I_3=+1$. By coupling it to the square of the Higgs field, we can obtain a gauge invariant lagrangian of the form:

\begin{equation}\label{dim5}
{\cal L}^{(5)}=\frac{y^2}{\Lambda} \bar{\ell}_L  \tilde{H}\tilde{H}^T  \ell_L^\star +h.c.
\end{equation}
where:
\begin{equation}
 \ell_L=\left(\begin{array}{c} \nu_L \\ e_L \end{array}\right)
 \label{e-doublet}
 \end{equation}
the Higgs doublet is organized as: 
 \begin{equation}
H=\left(\begin{array}{c}H^+ \\H^0 \end{array}\right), ~{\tilde H}_i=\epsilon_{ij} H_j=\left(\begin{array}{c}H^0 \\-H^+ \end{array}\right).
 \label{h-edoublet}
\end{equation}
and  we have introduced a numerical coupling  $y^2$. 

The term (\ref{dim5}) is non renormalizable because it has mass dimension $5$ ($3/2$ for each fermion field and
one for each Higgs field). For this reason, we have introduced a large mass in the denominator, $\Lambda$, of the order of the limit of validity of the Standard Theory, most naturally $\Lambda=M_{GUT}$.

\begin{figure}[!h]
        \centering
        {
       \includegraphics[scale=0.55]{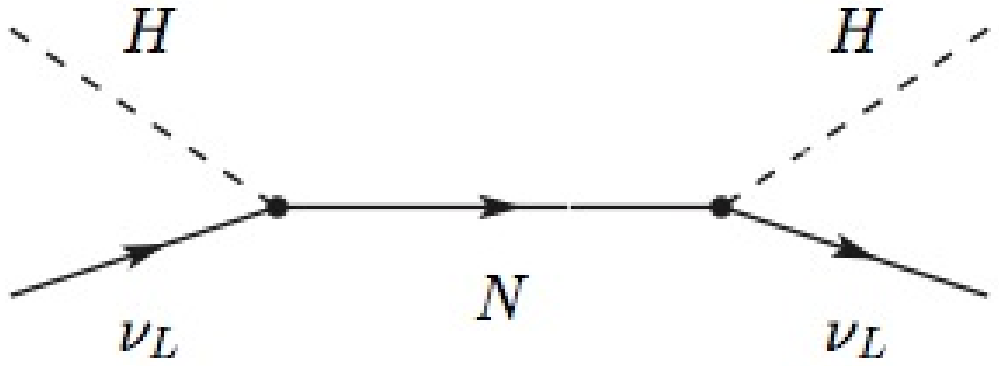}}
        {
           \includegraphics[scale=0.50]{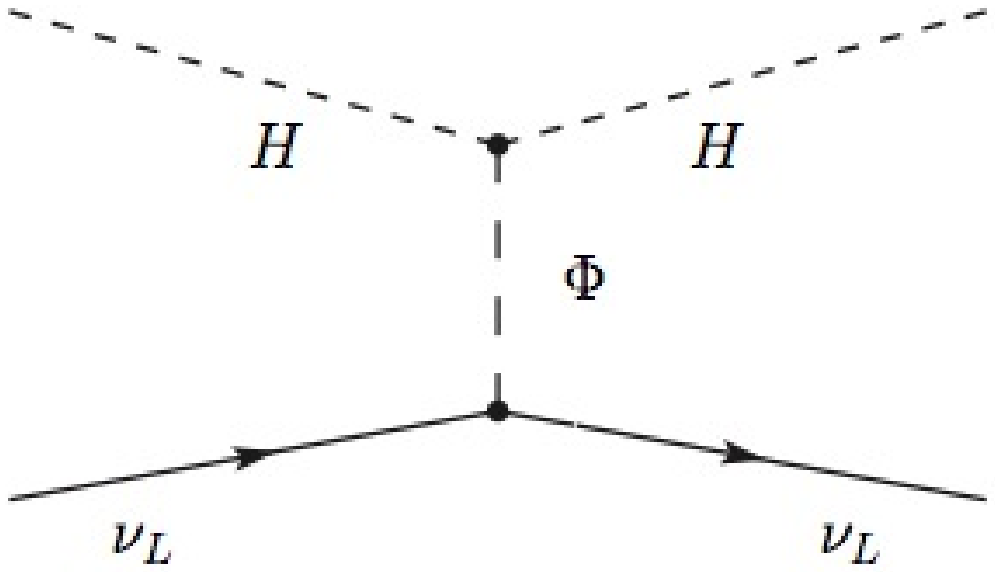}}
  	   \caption{\label{see-saw} {\small See-saw mechanism: (a) type I see-saw; (b) type II see-saw.}}
\end{figure}

When the Higgs field $H_2=\phi^0$ takes a vacuum expectaton value, $<0|\phi^0|0>=\eta$, the left-handed neutrino acquires a Majorana mass of the form given in (\ref{Mmass1}), with:
\begin{equation}
m_{\nu_L}=\frac{(y\eta)^2}{\Lambda}
\label{seesaw1}
\end{equation}
The similarity with the result (\ref{seesaw}) is evident.  In fact, we get exactly back to (\ref{seesaw}) if we  interpret the latter result as due to the exchange of a heavy fermion, as in Fig.~\ref{see-saw}(a). The mechanism we have just described is called Type I see-saw.

However, formula (\ref{seesaw1}) is more general, in that it could also be obtained by the exchange of a heavy scalar boson, as depicted in Fig.~\ref{see-saw}(b), which embodies Type II see-saw  mechanism.

A tiny Majorana neutrino mass could be the reflection of a more complicated theory at high energy which involves either heavy fermions or heavy scalars.

\subsection{\bf How can we tell a Majorana from a Dirac neutrino?} As we said before, $\beta$-decay and neutrino reactions have too large energy to allow detecting helicity/lepton number violating effects of order $(m_{\nu}/E_{\nu})^2$.
But there is a process where we can hope to understand if neutrinos are described by Majorana or by the Dirac-Weyl theory: the double beta decay without neutrinos ($\beta \beta \; 0 \nu$):

\begin{equation}\label{dbdecay}
X(A,Z)~ \rightarrow ~ X(A,Z \pm 2) + 2 e^{\mp},
\end{equation}

This process violates lepton number conservation by two units and it occurs if the neutrino coincides with its own antiparticles, i.e. is a Majorana particle.

For many years, the search for $\beta \beta \; 0 \nu$ has been a very specialized, almost marginal, matter.
Today it is an important research line of particle physics with experiments in the main underground laboratories around the world. In particular in the Gran Sasso Laboratory of INFN in Italy.

\begin{figure}[htbp]
        \centering
        \includegraphics[scale=0.4]{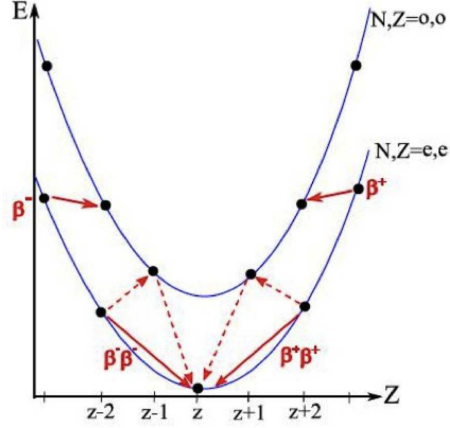}
        \caption{\label{massparab}{\small Mass parabola for isobaric nuclei, showing the necessary configuration for beta decay and double beta decay ($Z$ atomic number). 
There is $\beta^{-}$ decay when we start from a state $Z$ with more energy with respect to final state $Z+1$ so you gain energy by transforming a neutron 
into a proton. Conversely, $\beta^{+}$ decay occurs if you have more energy in the $Z+1$ state than in the state $Z$. Figure: courtesy of E. Fiorini. }}
\end{figure}

The possibility of double beta decay has been discussed first by Maria Goeppert-Mayer, in the thirties. We can illustrate these processes  with reference to Fig.~\ref{massparab}.

Usual $\beta^{-}$ decay occurs when a state $Z$ has more energy with respect to the  $Z+1$ state: you gain energy by transforming a neutron 
into a proton, as shown on the left side of the parabola in Fig.~\ref{massparab}. Conversely, $\beta^{+}$ decay occurs if the $Z+1$ state has more energy than the state  $Z$, as shown on the right side of the parabola in Fig.~\ref{massparab} . 

\begin{figure}[htbp]
       \centering
        \includegraphics[scale=0.3]{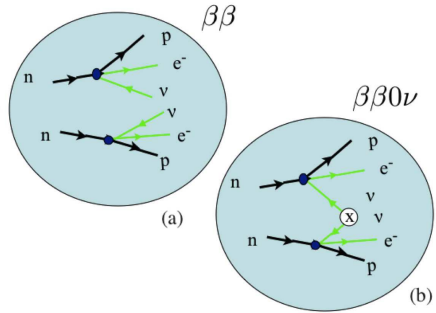}
        \caption{(a) Double beta decay; (b) Double beta decay without neutrinos. Figure: courtesy of E. Fiorini.}
        \label{doublebetadecay}
\end{figure}

However, there are cases of a contiguous triplet of states, where the energy of $Z$ is less than $Z+1$  and larger than $Z+2$ energy. In this case, $Z$ can decay directly in $Z+2$ at second order in the Fermi  interaction. This is double beta decay, which in turn may occur in two varieties: with or without neutrinos, illustrated in Figs.~\ref{doublebetadecay}.

\begin{figure}[htbp]
        \centering
       \includegraphics[scale=0.45]{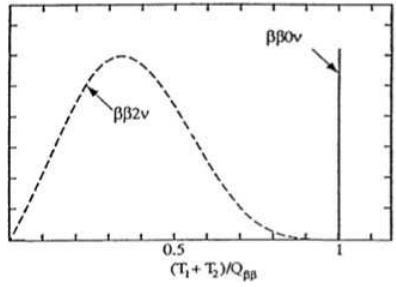}
  	   \caption{Energy spectrum for double beta decay and double beta decay without neutrinos. Figure: courtesy of E. Fiorini.}
  	   \label{energyspectrum}
\end{figure}

The process in Fig.\ref{doublebetadecay}-(a) is the normal double beta decay, in which there is a simultaneous transformation of two neutrons into two protons with an electron-neutrino pair emitted in each transition.  The presence of neutrinos  in the final states is indicated by the fact  that the sum of the energies of the two electrons observed in the final state shows a continuous spectrum (Fig.\ref{energyspectrum}), in correspondence to the unobserved energy carried by the neutrinos.

However, there is another possibility, Fig.\ref{doublebetadecay}-(b): if the two neutrinos are Majorana particles, the neutrino emitted by one neutron can be absorbed by the other neutron. In that case, the spectrum of the sum of the electrons energies is a perfect line (Fig.\ref{energyspectrum}). This is $\beta \beta 0\nu$ decay.

The process (\ref{dbdecay}) would not be possible for Dirac neutrino because of lepton number conservation while in Majorana theory this problem
disappears, replaced by a penalty due the neutrino mass, implied by the necessary helicity violation.

We can see this mathematically, starting from the current (\ref{betadecaycurrent}).

We need to compute the square of the current:
\begin{equation}
{\rm J}^\mu(x) {\rm J}^\nu(0)={\bar \psi_e} \gamma^\mu \frac{1}{2}(1-\gamma_5) <0|U(x)~U^T(0)\gamma^0|0>\frac{1}{2}
(1-\gamma_5)\gamma^\nu \psi_e^C,
\end{equation}
where the vacuum expectation value of the product $U(x)U^{T}(0)\gamma^0$ represents the exchange of the neutrino from one neutron to the other and:
\begin{equation}
<0|U(x)U^T(0)\gamma^0|0> = \frac{ \gamma^\mu k_\mu+m}{k^2 + m^2} ,
\end{equation}

Due to the $1-\gamma_5$ factors, the term proportional to $\gamma^\mu k_\mu$ drops and the amplitude, as we expected, is proportional to $m_{\nu}$. This  is coherent with the fact that in the limit of zero mass the Majorana theory is equivalent to Weyl theory, which implies lepton number conservation.

\section{{ Quark Masses and Mixing}}
\label{qm&m}

\subsection{\bf The quark flavor symmetry}

As we have seen, quarks appear in three generations, each consisting of a pair made by $Q=+2/3$, up-like, and $Q=-1/3$, down-like,  color triplets. 


Left-handed fields are arranged into doublets while right-handed fields are singlets under the gauge group $SU(2)_L\otimes U(1)_Y$ of the Weak and Electromagnetic interactions. We denote the doublets and singlets by: .
\begin{equation}
q_L=
\begin{pmatrix}
U_L \\
D_L
\end{pmatrix}
, \; \; U_R, \; \; D_R,
\end{equation}
where $U$ and $D$ indicate the up-type (up, charm and top) and down-type  (down, strange and bottom) quarks. Electro-weak and strong gauge interactions of quarks admit a large global symmetry:
\begin{equation}
{\cal G}_{quark}={\rm SU}(3)_{q}\otimes {\rm SU}(3)_{U}\otimes {\rm SU}(3)_{D} 
\label{gflav}
\end{equation}
and we have an $SU(3)$ group for each generation triplet, $q$, $U_R$ and $D_R$, respectively. This is the {\it{\bf quark flavor symmetry}}.

Flavor symmetry is {\it explicitly broken} by the Yukawa couplings of quarks to the Higgs doublet, to avoid the proliferation of unobserved Goldstone bosons (we shall reconsider this point in Sect.~\ref{dynyuk}).

Couplings are encoded in complex, $3\times 3$ numerical matrices in the space of the generations:
\begin{equation}
 {\cal L}_Y={\bar q}_L {\it Y}_D H D_R +{\bar q}_L {\it Y}_U {\tilde H} U_R~+~{\rm h.c.}, \label{ylag}
\end{equation}
and the Higgs fields are organized as in eq.~(\ref{h-edoublet}):
\begin{equation}
H=\left(\begin{array}{c}H^+ \\H^0 \end{array}\right), ~{\tilde H}_i=\epsilon_{ij} H_j=\left(\begin{array}{c}H^0 \\-H^+ \end{array}\right).
\end{equation}

The gauge symmetry is broken by the vacuum expectation value of the neutral component:
\begin{equation}
\langle0|H^0|0\rangle=\eta
\label{vev}
\end{equation}

Replacing the vacuum value (\ref{vev}) in the Yukawa lagrangian~(\ref{ylag}) one finds the quark mass matrices:
\begin{equation}
M_D= {\it Y}_D \eta~, ~ M_U={\it Y}_U \eta, \label{mass term}
\end{equation} 

Similarly to the ${\it Y}$s, $M_{D,U}$ are $3\times 3$ matrices in generation space, complex matrices if the ${\it Y}$s are complex.

Yukawa couplings can be diagonalized by bi-unitary transformations by virtue of a simple theorem in matrix theory and its immediate corollary, see Appendix I.
\begin{itemize}
\item {\bf Theorem.} Any complex matrix $Y$ can be written as
\begin{equation} 
Y= HW 
\label{theorem}
\end{equation}
with $W$ unitary and $H$ hermitian and non negative;
\item {\bf Corollary}. Any complex matrix $Y$ can be written as 
\begin{equation} 
Y=U \rho V 
\label{corollary}
\end{equation}
 with $U$ and $V$ unitary and $\rho$ diagonal, with real, positive or zero, elements.
\end{itemize}

\paragraph{\bf Note} The transformations of the flavor group, (\ref{gflav}), being of unit determinant, cannot change an overall phase in $Y$. So to have $\rho$ real, we must either assume the $Y$ has real determinant or that we can multiply additional $U(1)$ factors in ${\cal G}_{quark}$. The latter transformations are in general anomalous and call into play the anomalous QCD parity-violating lagrangian $G^{\mu \nu}\tilde G_{\mu \nu}$. Introducing a field associated to the phase of $Y$ may eliminate the problem of parity-violation in strong interactions, the so-called $\theta$ puzzle, as pointed out originally by A. Peccei and H. Quinn\cite{Peccei:1977hh}
at the cost of introducing a new particle, the axion. For a recent discussion, see e.g. Ref.~\cite{Fong:2013sba} and references therein. We shall avoid getting into all the subsequent complications and assume that $Y$ is, or can be made, real and all unitary matrices needed to diagonalize the $Y$s have unit determinant.

\subsection{\bf Quark masses and mixing}

Using the corollary mentioned above, we can write the Yukawa couplings in (\ref{ylag}) as:
\begin{equation}\label{YuYd}
Y_{U}=W \rho_{U} Z, \;\;\;\; Y_{D}=U \rho_{D} V
\end{equation}

The lagrangian in (\ref{ylag}), with $H\to \langle0|H|0\rangle$ and $\eta~\rho_{U,D}=m_{U,D}$, becomes:
\begin{equation}
 {\cal L}_Y={\bar D}_L Um_{D}V D_R +{\bar U}_L Wm_{U} Z U_R~+~{\rm h.c.}, \label{ylagd}
\end{equation}

Without breaking the gauge symmetry, we are free to redefine the singlets and the doublets with unitary tranformations, that is, to set
\begin{equation}
V D_R\to D_R,~ZU_R\to U_R,~W^\dagger q_L\to q_L
\end{equation}
and the mass quark Lagrangian becomes:
\begin{equation}
{\cal L}_{mass}= \bar{D}_{L} W^\dagger U m_{D} D_{R}+ \bar{U}_{R} m_{U} U_{L},
\end{equation}

With this choice, up quark fields are at the same time mass and weak isospin eigenstates, while for down quarks we have to make one further redefiniton which breaks the $SU(2)_L$ symmetry:
\begin{equation}
D^{\dagger}_{weak,L}(W^\dagger U)=D_{mass,L}^\dagger,~i.e.~D_{weak,L}=(W^\dagger U) D_{mass,L}=U_{CKM}D_{mass,L}
\label{ckmdef}
\end{equation}

The mass lagrangien, in terms of the new fields is completely diagonal. However the weak interactions couple the weak fields so that, expressing the latter in terms of fields which are the eigenstates of the masses, via eq.~(\ref{ckmdef}), we obtain the weak coupling to the $SU(2)_L$ vector bosons in the form:
\begin{eqnarray}
&&{\cal L}_{weak}=\sqrt{2}~{\bar U}\gamma^\mu (1-\gamma_5)U_{CKM}D W_\mu~+ ~{\rm h.c.}~+ \notag \\
&&+\left[{\bar U}\gamma^\mu (1-\gamma_5)U-{\bar D}(U_{CKM}^\dagger\gamma^\mu (1-\gamma_5)U_{CKM})D\right]W^3_\mu
\label{charcurr}
\end{eqnarray}
which identifies $U_{CKM}$ with the Cabibbo-Kobayashi-Maskawa matrix~\cite{Ncab,KM}, responsible for flavor violation in quark decays. There is no flavor violation in the neutral current interaction, since $U_{CKM}$ is unitary~\cite{gim}.

With three generations,  the CKM matrix depends upon three real parameters and one CP violating phase. In the very convenient parametrization due to Wolfenstein~\cite{wolfckm} one has: 
\begin{equation}
U_{CKM} =
\begin{pmatrix}
 1-\frac{1}{2}\lambda^2 & \lambda &A\lambda^3(\rho-i\eta)\\  &   &   \\-\lambda & 1-\frac{1}{2}\lambda^2 & A\lambda^2 \\ &   &   \\A\lambda^3[1-(\rho+i\eta)] & -A\lambda^2  & 1
\end{pmatrix}
.
\end{equation}
with $\lambda = \sin \theta_C$, $\theta_C$ being the Cabibbo angle.

Parameters of the CKM matrix have been determined experimentally to a good accuracy, see~\cite{Ceccucci:2008zz}
\begin{eqnarray}
\lambda = 0.2253 \pm 0.0007,\; \; A= 0.808^{+0.022}_{-0.015},  \nonumber \\
  \bar \rho = 0.132^{+0.022}_{-0.014}, \;\ \;  \bar \eta = 0.341 \pm 0.013,
\end{eqnarray}
with:
\begin{equation}
\bar \rho+i\bar \eta =(\rho+i\eta)(1-\frac{\lambda^2}{2}) + {\cal O}(\lambda^4)
\end{equation}

Numerically
\begin{equation}
U_{CKM} =
\begin{pmatrix}
0.9746 & 0.2253 & 0.0012- i 0.0032 \\
-0.2253 & 0.9746 & 0.0410\\ 
0.0080 - i 0.0032  & -0.0410 & 1
\end{pmatrix}.
\end{equation}

The CKM matrix is close to the unit matrix, with non diagonal elements decreasing approximately with powers of the small parameter $\lambda$. 

\begin{figure}[!ht]
        \begin{center}
       \includegraphics[scale=0.5]{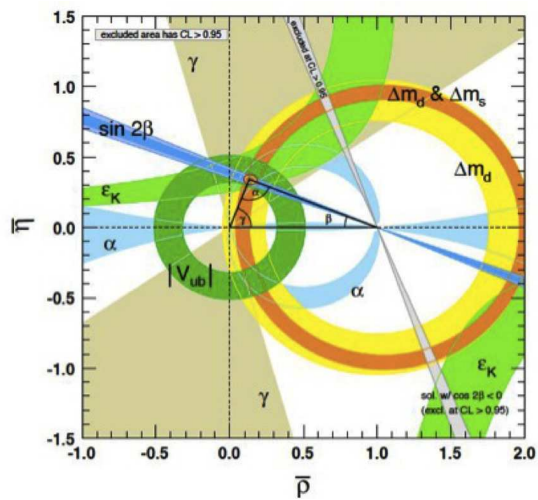}
  	   \caption{\label{FittingCKM} \small Constraints in the $\bar{\rho}$, $\bar{\eta}$ plane. Superimposed are the individual constraints from charmless semileptonic $B$ decays ($|V_{ub}|$), mass differences in the $B_d$ and $B_s$ mesons ($\Delta m_d$ and $\Delta m_s$), CP violation in the neutral K meson ($\epsilon_K$) and in the $B_d$ systems ($sin 2\beta$). Figure from Ref.~\cite{Ceccucci:2008zz}.
}   
\end{center}
\end{figure}

The quality of the present determination of  the CP violating parameter, $\bar \rho+i\bar \eta$, is remarkable. Constraints arising from different weak interaction observables are shown in Fig.~\ref{FittingCKM}, taken from Ref.~\cite{Ceccucci:2008zz}.

\section{ FCNC Processes: Standard Theory and Beyond}
\label{fcnc}
In the ST, we have a good control on a number of  {\it flavor changing neutral current} effects (FCNC). 

The typical case, also the first that has been studied, is $K^{0}$ -$\bar{K}^{0}$ mixing, which gives rise to the $K_L-K_S$ mass difference and to the observed CP violation in $K_L$ decay. 

After computing the effective Lagrangian for this transition, we end up with the expression for the off-diagonal element of the mass matrix (see e.g.~\cite{burasetal} and references therein):

\begin{equation}\begin{split}
&M_{12}(\bar K^0\to K^0) =  < K^0|-{\cal L}_{eff}|\bar K^0>= \\
&=\frac{(G_F M_W^2)(G_F f_K^2)}{12\pi^2}  \times \sum_{i,j=c,t}C_i C_j E(x_i, x_j)\times m_K, \label{M12}
\end{split}\end{equation}
$E(x,y)$ are the so-called Inami-Lim~\cite{inamilim} loop factors, with $x=(m_{q}/M_{W})^2$ , and the combination of the CKM coefficients $C_i C_j$ are:. 
\begin{equation}
C_iC_j=(U_{id}U^*_{js})(U_{id}U^*_{js});\;(i, j=c, u)
\end{equation}
We can also add QCD corrections represented by some computable coefficients $\eta_1$, $\eta_2$ and $\eta_3$ (see ~\cite{ciuchini} and references therein) so that
\begin{equation}\begin{split}
&M_{12}(\bar K^0\to K^0)|_{corr}=\frac{(G_F M_W^2)(G_F f_K^2)}{12\pi^2}  \times \\
&\times \left[ \eta_1 C_c^2E(x_c, x_c)   +   \eta_2 C_t^2 E(x_t, x_t) +2 \eta_3 C_c C_t E(x_c, x_t)\right]\times m_K  \times B_K , \label{M12 QCD}
\end{split}\end{equation}
$B_K$ being the $B$-factor that takes into account the intermediate states following the vacuum state.

Transitions which change flavor by two units ($\Delta F =2$) for $K^0$ mesons are dominated by c quark and, to a lesser extent,  by t quark, so that we may trust the values of the QCD coefficients computed with improved perturbation theory. 

On the other hand, $\Delta F =2$ transitions for D-mesons are dominated by s and b quarks, but b quark exchange is CKM suppressed much more than s. We obtain
\begin{equation}
M_{12}(\bar D^0\to D^0)=\frac{(G_F M_W^2)(G_F f_D^2)}{12\pi^2}\times \sum_{i,j=s,b} C_iC_j E(x_i, x_j) \times m_D , \label{M12 total}
\end{equation} 
with $C_b\approx (\sin\theta_C)^5$ and $C_s\approx (\sin\theta_C) $. 

Finally, transitions for B-mesons correspond to processes dominated by t quark and the QCD corrected matrix element for these transitions is 
given by
\begin{equation}
M_{12}(\bar B^0\to B^0)|_{corr}=\frac{(G_F M_W^2)(G_F B_B f_B^2)}{12\pi^2}\times \eta_b C_t^2 E(x_t, x_t) \times m_B ,
\end{equation}
where $\eta_b=0.55$ represents the QCD correction, $B_B$ is the appropriate $B$-factor and $f_B$ is the decay constant for $B$ meson 
transitions.

Table~\ref{confronto_dati} shows the comparison with experimental data, before and after introducing QCD corrections. The only case of disagreement is the one which is dominated by strange quark, i.e. a low-mass quark, in a range where perturbative QCD is not reliable~\footnote{It is worth noting that the CKM coefficients for $K^0-\bar K^0$ mixing in (\ref{M12 QCD}) are such as to suppress greatly the $top$ quark exchange diagrams in the real part of $M_{12}$. Therefore, the connection of the $charm$ quark mass to the $K_L-K_S$ mass difference, pointed out originally in Ref.~\cite{gim} in the four-quark scheme, remains valid in the CKM, six-quark, scheme as well.}.
\begin{table}[htb]
\begin{center}
\vskip0.3cm
\tabcolsep=0.08mm
\begin{tabular}{@{}ccccccc}
\hline 
      & $|\epsilon_K|$  &  $\Delta m_K$  &  $|\Delta M(B^0_d)|$  &    $|\Delta M(B^0_s)|$& $|\Delta M(D^0)|$&$Br(B_s \to \mu^+\mu^-)$ \\
  \hline
            &     &     &   &  & & \\           
EW & {\small $6.34\;10^{-3}$  } & {\small $3.12\;10^{-12}$ } & {\small $7.51\;10^{-10}$}  &  {\small $2.94\;10^{-8}$}&{\small $2.0\;10^{-13}(\frac{m_s}{0.15 GeV})^2$}  &{\small $4.0\;10^{-9}$ } \\ 
\\
 QCD corr &  {\small $2.65\;10^{-3}$ }   &  {\small $3.85\;10^{-12}$ }   &   {\small $4.13\;10^{-10}$ }  &  {\small $1.19\;10^{-8}$ }& {\small not reliable }&{\small $(3.53\pm0.38)\;10^{-9}$ } \\
\hline
            &     &     &     & & \\ 
            expt  &  {\small $2.228\;10^{-3}$ }    &   {\small $3.483\;10^{-12}$ } &   {\small $3.34\;10^{-10}$ }  &  {\small $1.17\;10^{-8}$ }&$(1.57\pm0.39)\;10^{-11}$ & {\small $(3.2\pm1.4)\;10^{-9}$ } \\
 \hline
\end{tabular}\\[2pt]
\caption{ {\footnotesize {\rm Comparison of ST predictions with data for some FCNC processes. The observables considered are: $\epsilon_K$,  the CP violation amplitude for K meson, $\Delta m_K $, the $K_L-K_S$ mass difference, $\Delta M(B)$  and $\Delta M(D)$), the mass differences between the heavy and light eigenstates of the $B$ and $D$ neutral mesons and the branching ratio for the decay $B_s \to \mu^+ \mu^-$~\cite{bsmumudec}. Masses in MeV. Table from Ref.~\cite{electroweak}.}}}
\label{confronto_dati}
\end{center}
\end{table}

\subsection{\bf GIM mechanism and limits on the scale of new physics}

The good agreement with experimental data of FCNC can be used to give limits to the energy scale of effects beyond the ST. 

There have been suggestions that new physics (NP) may exist,  related to SuperSymmetry (SUSY) at TeV scale. SUSY particles 
should of course carry flavor and contribute to $K^{0}$- $\bar{K}^{0}$ mixing.  The latter contributions take at low energy the form a general local Lagrangian which includes operators with dimension $d =6$, constructed in terms of ST fields, and suppressed by inverse powers of an effective scale $\Lambda$, which characterizes the scale of NP.

We write the total effective lagrangian as, e.g.: 
\begin{equation}\begin{split}
&{\cal L}_{eff}(d \bar s\to \bar d s) = {\cal L}_{ST}+{\cal L}_{NP}=  \\
&= -\frac{G_F^2 M_W^2}{16\pi^2}  \times \sum_{i,j=c,t}(U^*_{id}U_{is})( U^*_{jd}U_{js}) E(x_i, x_j)\times \left[\bar d_s\gamma^\mu(1-\gamma_5)d\right]\left[\bar d_s\gamma_\mu(1-\gamma_5)d\right]+ \\
& +\frac{c_\Gamma}{\Lambda^2} \left(\bar d\Gamma^k  s\right)\left(\bar d \Gamma^k s\right) \label{L eff}
\end{split}\end{equation}
where the first term is the ST contribution and the second term the NP contribution, for a dimension six operator determined by some four fermion covariant.
Since ${\cal L}_{ST}$ reproduces well the data, we must require $|NP| < |ST|$ and we obtain limits that we can organize in two ways:
\begin{itemize}
\item assume $c_\Gamma \sim 1$ and obtain a limit on $\Lambda$,
\item assume $\Lambda\sim 1$~TeV and obtain a limit on $|c_\Gamma|$.
\end{itemize}

We insert one dimension $d =6$ operator at a time, assuming there are no cancellations among the NP amplitudes, and obtain the results summarized in Tab. \ref{tabGIM}, Ref.~\cite{Isidori2012}.
\begin{table}[h]
\begin{center}
\begin{tabular}{cccccc} \hline
\rule{0pt}{1.2em}%
Operator &  \multicolumn{2}{c}{Bounds on $\Lambda$~in~TeV~($c_{\rm NP}=1$)} &
\multicolumn{2}{c}{Bounds on
$c_{\rm NP}$~($\Lambda=1$~TeV) }& Observables\cr
&   Re& Im & Re & Im \cr  
 \hline $(\bar s_L \gamma^\mu d_L )^2$  &~$9.8 \times 10^{2}$& $1.6 \times 10^{4}$ 
&$9.0 \times 10^{-7}$& $3.4 \times 10^{-9}$ & $\Delta m_K$; $\epsilon_K$ \\ 
($\bar s_R\, d_L)(\bar s_L d_R$)   & $1.8 \times 10^{4}$& $3.2 \times 10^{5}$ 
&$6.9 \times 10^{-9}$& $2.6 \times 10^{-11}$ &  $\Delta m_K$; $\epsilon_K$ \\ 
 \hline $(\bar c_L \gamma^\mu u_L )^2$  &$1.2 \times 10^{3}$& $2.9 \times 10^{3}$ 
&$5.6 \times 10^{-7}$& $1.0 \times 10^{-7}$ & $\Delta m_D$; $|q/p|, \phi_D$ \\ 
($\bar c_R\, u_L)(\bar c_L u_R$)   & $6.2 \times 10^{3}$& $1.5 \times 10^{4}$ 
&$5.7 \times 10^{-8}$& $1.1 \times 10^{-8}$ &  $\Delta m_D$; $|q/p|, \phi_D$\\ 
\hline$(\bar b_L \gamma^\mu d_L )^2$    &  $6.6 \times 10^{2}$ & $ 9.3 \times 10^{2}$ 
&  $2.3 \times 10^{-6}$ &
$1.1 \times 10^{-6}$ & $\Delta m_{B_d}$; $S_{\psi K_S}$  \\ 
($\bar b_R\, d_L)(\bar b_L d_R)$  &   $  2.5 \times 10^{3}$ & $ 3.6
\times 10^{3}$ &  $ 3.9 \times 10^{-7}$ &   $ 1.9 \times 10^{-7}$ 
&   $\Delta m_{B_d}$; $S_{\psi K_S}$ \\
\hline $(\bar b_L \gamma^\mu s_L )^2$    &  $1.4 \times 10^{2}$ &  $  2.5 \times 10^{2}$   &  
 $5.0 \times 10^{-5}$ &   $1.7 \times 10^{-5}$ 
   & $\Delta m_{B_s}$; $S_{\psi \phi}$ \\ 
($\bar b_R \,s_L)(\bar b_L s_R)$  &    $ 4.8  \times 10^{2}$ &  $ 8.3  \times 10^{2}$  & 
   $8.8 \times 10^{-6}$ &   $2.9 \times 10^{-6}$  
  & $\Delta m_{B_s}$;  $S_{\psi \phi}$ \\ \hline
\end{tabular}
\caption{\label{tabGIM} {\rm Bounds on representative $d= 6$, $\Delta F=2$  operators, assuming an effective coupling $c_{\rm NP}/\Lambda^2$. The bounds quoted are: (i)  on $\Lambda$, setting $|c_{\rm NP}|=1$, (ii) on  $c_{\rm NP}$, setting
$\Lambda=1$ TeV. In the right column the main observables used to derive these bounds. Table from Ref.~\cite{Isidori2012}. }}
\end{center}
\end{table}


It is clear that if NP exists at all at the TeV scale, it cannot be coupled {\it generically} to flavor.

Several proposals have been advanced, started from the seminal paper by Chivukula and Georgi~\cite{georgi}. Most developed is the proposal advanced in~\cite{mfv}, the {\it {\bf Minimal Flavor Violation Principle}}, MFV, summarized by the statement that: {\it Yukawa couplings are the only source of flavor symmetry violation, even for 
NP}.

\subsection{\bf Minimal Flavor Violation}
We can introduce the idea of MFV by first noting that the Lagrangian (\ref{ylag}) is not invariant under the flavor group ${\cal 
G}_{quark}=SU(3)_q\otimes SU(3)_{U_R}\otimes SU(3)_{D_R}$ because $Y$s couplings are fixed numbers. However 
(\ref{ylag}) would be invariant under the flavor group if  the Yukawa couplings would be subjected to the same ${\cal G}_{quark}$ 
transformations as quark fields:
\begin{equation}
q_L \to U_L q_L; \;  D_R \to V D_R; \; \; U_R\to W U_R, \label{field transf}
\end{equation}
\begin{equation}
{\it Y}_D \to U_L {\it Y}_D V^\dagger; \; \;  {\it Y}_U \to U_L {\it Y}_U W^\dagger, \label{Y transfo}
\end{equation}
$U_L$, $V$ and $W$ being unitary matrices. Formal transformations of the lagrangian coefficients have been used in the past to find the selection rules of the effect of some symmetry breaking. The term``spurion' was used to indicate $Y$s that formally transform like fields.

Assume now that NP is made of particles transforming non trivially under ${\cal 
G}_{quark}$, with masses of order $\Lambda$, much larger than the electroweak scale. Assume further that the symmetry breaking in the new sector is described by the same Yukawa couplings, $Y_D$ and $Y_U$ in such a 
way as to make also the Lagrangian for the new particles to be invariant under the combined transformations (\ref{field transf}) and (\ref{Y 
transfo}), supplemented with the transformations of the new particles. 

 FCNC effects produced by NP will give rise, at the electroweak scale, to effective lagrangians described by dimension $d=6$, four fermion operators, such as, e.g. 
\begin{equation}\label{fermionicoperator}
\frac{1}{\Lambda^2}\left[{\bar d}\gamma^\mu(1-\gamma_5) s\right]\cdot\left[{\bar d}\gamma_\mu(1-\gamma_5) s\right],
\end{equation}
but the coefficient must contain appropriate powers of $Y_D$ and $Y_U$ so as {\it to make the overall operator invariant under transformations of fields and spurions}.

the principle of Minimal Flavor Violation implies that the couplings appearing in front of effective NP operators, will be suppressed by CKM angles in a way similar to what happens for the ST effective lagrangian. The effect is to release considerably the bounds on $\Lambda$. 

For example, to satisfy MFV, the operator (\ref{fermionicoperator}) has to be part of the spurion-containing invariant operator:

\begin{equation}
\dfrac{1}{\Lambda^2} \left( \bar{q}_L Y_U Y_{U}^{\dagger} \gamma_{\mu} q_L \right)^2.
\label{invariant}
\end{equation}
This expression has coefficients coming from $Y_U Y_U^\dagger$ which include the very small angles that appear in the CKM matrix and so it is naturally suppressed
\footnote{the down field appearing in $q_L$ is $ D_L=U_{CKM} D_{mass, L}$, see eq. (\ref{ckmdef}), so that the term containing $D$ fields in eq. (\ref{invariant}) are obtained with substitution: $Y_U Y_U^\dagger \to  U_{CKM}^\dagger Y_U Y_ U^\dagger  U_{CKM}$ and the relevant amplitude for the $\Delta S=2$  transition contains a factor: $[(U_{CKM}^\star)_{13}(U_{CKM})_{23}]^2 m_t^4$ , mimicking the CKM factors appearing in the ST amplitude.}. 

We can see the effect of MFV in Table \ref{MFV}, which shows  the bounds on $\Lambda$, for each four-fermion operator that may be produced by NP, Ref.~\cite{Isidori2012}. 
\begin{table}[h]
\begin{center}
\begin{tabular}{lcl}
Operator & ~Bound on $\Lambda$~  & ~Observables \\
\hline
$H^{\dagger} \left( \bar{D}_R Y_d^\dagger  Y_u Y_u^\dagger
  \sigma_{\mu\nu} Q_L \right) (e F_{\mu\nu})$ 
& ~$6.1$~TeV & ~$B\to X_s \gamma$, $B\to X_s \ell^+ \ell^-$\\
$  \frac{1}{2} (\bar{Q}_L  Y_u Y_u^\dagger \gamma_{\mu} Q_L)^2  \phantom{ \Big( }$  
& ~$5.9$~TeV & ~$\epsilon_K$, $\Delta m_{B_d}$, $\Delta m_{B_s}$ \\   
$H_D^\dagger \left( \bar{D}_R  Y_d^\dagger 
Y_u Y_u^\dagger  \sigma_{\mu\nu}  T^a  Q_L \right) (g_s G^a_{\mu\nu})$
&~$3.4$~TeV & ~$B\to X_s \gamma$, $B\to X_s \ell^+ \ell^-$\\
$\left( \bar{Q}_L Y_u Y_u^\dagger  \gamma_\mu
Q_L \right) (\bar{E}_R \gamma_\mu E_R)$  
& ~$2.7$~TeV &~$B_s\to\mu^+\mu^-$,~$B\to X_s \ell^+ \ell^-$  \\
$~i \left( \bar{Q}_L Y_u Y_u^\dagger  \gamma_\mu
Q_L \right) H_U^\dagger D_\mu H_U$
&~$2.3$~TeV 
&~$B_s\to\mu^+\mu^-$,~$B\to X_s \ell^+ \ell^-$\\
$\left( \bar{Q}_L Y_u Y_u^\dagger  \gamma_\mu Q_L \right) 
( \bar{L}_L \gamma_\mu L_L)$
&~$1.5$~TeV &~$B_s\to\mu^+\mu^-$,~$B\to X_s \ell^+ \ell^-$\\
$\left( \bar{Q}_L Y_u Y_u^\dagger  \gamma_\mu Q_L
\right) (e D_\mu F_{\mu\nu})$
&~$1.7$~TeV & ~$B\to X_s \ell^+ \ell^-$\\
\hline
\end{tabular}
\end{center}
\caption{\label{MFV} {\rm Bounds on the scale of new physics (at 95\%
  C.L.) for some representative  MFV operators, assuming effective coupling $\pm 1/\Lambda^2$, and considering only one operator at a  time.  
The observables used to set the bounds are indicated in the last column. Table from Ref.~\cite{Isidori2012}.}}
\end{table}

\begin{figure}[h]
        \centering
       \includegraphics[scale=0.4]{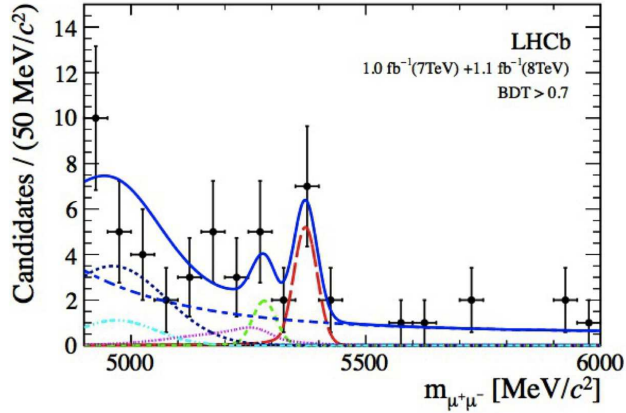}
  	   \caption{  \label{LHCb} Evidence of the $B_s^0 \rightarrow \mu^+ \mu^-$ decay~\cite{bsmumudec}. The black dots represent the number of events detected in each energy range while the blue line represent the theoretical prediction of the ST. Figure from Ref.~\cite{bsmumudec}.}
\end{figure}

The situation is even more comfortable if NP produces FCNC effects in low energy processes by loop diagrams. In this case, rather than $c_\Gamma \sim 1$, one would have $c_\Gamma \sim g^2/4\pi$ with some coupling constant $g$. A good example is the Constrained Minimal Supersymmetric Standard Model (CMSSM), which is a SUSY model which satisfies MFV and which shows that the limits from FCNC are compatible with a relatively 
low energy scale for NP. This is the case of the rare decay $B_s \rightarrow \mu^+ \mu^-$ , observed recently by the LHCb collaboration~\cite{bsmumudec}, see Fig.~\ref{LHCb}, with the branching ratio:
\begin{equation}
{\cal B}(B_s^0 \rightarrow \mu^+ \mu^-)=(3.2^{+1.4}_{-1.2} (stat)^{+0.5}_{-0.3} (syst)) \times 10^{-9},
\label{bsmu}
\end{equation}
very close to the ST prediction, reported in Tab.~\ref{confronto_dati}.
	
\begin{figure}[h]
        \centering
       \includegraphics[scale=0.4]{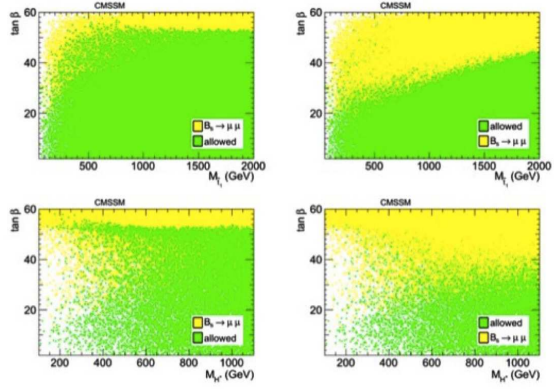}
  	   \caption{ \label{Constraint_CMSSM} Constraint from ${\cal B}(B_s \rightarrow \mu^{+} \mu^{-})$ in the CMSSM plane ($M_{\tilde{t}_1} , \tan\beta$) in the upper
panel and ($M_{H^{\pm}} , \tan \beta$) in the lower panel, with the allowed points displayed in the foreground in the left and in the background in the right ($\tilde{t}$ is the scalar corresponding to the $top$ quark and $\tan\beta$ is the ratio of the vacuum expectation values of the two Higgs doublets implied by Supersymmetry). Figure taken from Ref.~\cite{mahmoudi}.}
\end{figure}

 The measurement of the decay (\ref{bsmu}) provides important constraints on the masses of new particles predicted by SUSY, namely: $\tilde t$, the scalar partner  of the $top$ quark, and $H^\pm$, the charged Higgs bosons, see Fig \ref{Constraint_CMSSM} taken from Ref.~\cite{mahmoudi}. Limits are still compatible with the negative results of present SUSY searches at the LHC. 
\subsection{{\bf MFV versus quark masses}}

The MFV principle can be illustrated by the familiar case of the violation of chiral symmetry in QCD. 

Quark masses are {\it the sole source of chiral symmetry violation} and the QCD lagrangian is invariant under chiral symmetry if we treat quark masses as spurions. At low energy, chiral symmetry violating hadronic processes are described by effective lagrangians whose coefficients must depend upon the quark masses, always the same in all processes, in such a way as to be chiral invariant if we transform hadron fields and quark masss at the same time.

In the early times this was the hypothesis of ($ 3, \bar 3$) transformation of the symmetry breaking lagrangian under chiral $SU(3)\otimes SU(3)$. The coefficients of the ($3,\bar 3$) operator have been  later interpreted as the quark masses which determine universally the breaking. 

Similarly, with respect to the flavor group (\ref{gflav}), the Yukawa spurions would transform as:
\begin{equation}
Y_U\sim (3,\bar 3,1);~Y_D\sim(3,1,\bar 3)
\label{spurtransf}
\end{equation}
which show that the expression (\ref{invariant}) is indeed invariant under combined transformations of fields and spurions. Spurions provide an efficient bookkeeping of the predicitions of symmetry breaking for the coefficients of the effective lagrangians.

Similarly to QCD, the Yukawa couplings have been interpreted in~\cite{georgi} as a consequence of flavor breaking {\it preon masses} (preons being the supposed elementary constituents of quarks and leptons). We shall see later, Sect.~\ref{dynyuk}, a different interpretation of the universality of the Yukawa couplings implied by MFV.

\newpage

\section{{ Neutrino Mixing and Oscillations with Three Flavors}}
\label{num&m}

Now we turn our attention to the argument of lepton flavor, in particular neutrino mixing and oscillations. Neutrino mixing and oscillations have been introduced by Bruno Pontecorvo~\cite{pontecorvo,Bilenky:2013wna} and by Shoichi Sakata and Collaborators~\cite{makietal}, considering the case of two neutrinos. Neutrino oscillations with 3 flavours including CP and CPT violation was discussed  by  Cabibbo~\cite{cabnu} and by Bilenky and Pontecorvo~\cite{Bilenky:1978nj}.

We can obtain the neutrino mixing matrix with the same argument that led us to the CKM matrix. If we set ourselves in the field basis where charged leptons are diagonally flavoured, the mixing of the three generations of neutrinos is described by a $3 \times 3$ complex matrix, known as the Pontecorvo-Maki-Nakagawa-Sakata (PMNS) mixing matrix, determined by three real angles and one CP violating phase.

We write the weak current according to:
\begin{equation}
{\it J}^\mu = \left(\bar{e},\bar{\mu},\bar{\tau} \right) \gamma^{\mu} \left(1-\gamma_5 \right)V_{PMNS}
\begin{pmatrix}
\nu_1 \\
\nu_2 \\
\nu_3
\end{pmatrix},
\end{equation}

The latest data determine the three real angles, but we do not have yet information on the CP violating phase. Numerically:
\begin{equation}
V_{PMNS} \approx
\begin{pmatrix}
0.822 & 0.549 & 0.15 e^{i\delta} \\
-0.394 +0.084 e^{-i\delta}& 0.591+ 0.069 e^{ -i\delta} & -0.653 \\ 
0.367 + 0.090 e^{-i\delta}  & -0.550 + 0.060 e^{-i\delta} & -0.702
\end{pmatrix}
\label{pmatrix}
\end{equation}

Unlike CKM, the PMNS matrix, has large non-diagonal elements. Quite a surprise. 

\subsection{\bf Deriving the mixing matrix}To derive the  PMNS matrix we proceed as follows. 

We treat neutrinos as Weyl particles (we will see in Sect.~\ref{dynyuk} that the result is correct also for see-saw Majorana neutrinos) and write first the field $\nu_e$, the field coupled in the weak current,  in the basis of the fields $\nu_{1,2,3}$, which diagonalize the mass matrix. In general: 
\begin{equation}
\nu_e=\cos \theta_{13}\left[\cos\theta_{12}\nu_1+\sin\theta_{12}\nu_2\right]+e^{i\delta}\sin\theta_{13}\nu_3. \label{nu_e}
\end{equation}
After that, we define further two orthonormal fields with respect to $\nu_e$
\begin{equation}
\nu^\prime=-\sin\theta_{12}\nu_1+\cos\theta_{12}\nu_2, 
\end{equation}
\begin{equation}
\nu^{\prime \prime}= e^{-i\delta}\sin\theta_{13}\left[\cos\theta_{12}\nu_1+\sin\theta_{12}\nu_2\right]-\cos\theta_{13}\nu_3,
\end{equation}
and the angle $\theta_{23}$ is defined by
\begin{equation}
\nu_\mu=-\cos\theta_{23}\nu^\prime+\sin\theta_{23}\nu^{\prime \prime},
\end{equation}
\begin{equation}
\nu_\tau= -\sin\theta_{23}\nu^\prime+\cos\theta_{23}\nu^{\prime \prime}. \label{nu_tau}
\end{equation}

Relations (\ref{nu_e})--(\ref{nu_tau}), give us the most general form of the PMNS matrix
\begin{equation} \label{general PMNS}
V_{PMNS}  =
\begin{pmatrix}
c_{13}c_{12} & c_{13}s_{12} & e^{i\delta}s_{13}\\
-c_{23}s_{12}+e^{-i\delta}s_{23}s_{13}c_{12} & c_{23}c_{12}+e^{-i\delta}s_{23}s_{13}s_{12} & -s_{23} c_{13}\\
s_{23}s_{12}+ e^{-i\delta}c_{23} s_{13} c_{12} & -s_{23}c_{12}+ e^{-i\delta} c_{23}s_{13}s_{12} & -c_{23}c_{13} 
\end{pmatrix}
\end{equation}

\subsection{\bf Neutrino oscillations} 

The amplitude $A$ 
for the appearence of flavor $j$ neutrino at a distance $L$ from the production of flavor $i$ neutrino is:
\begin{eqnarray}
&& A(i \to j,L)=\sum_{a, b}\; \langle j|a\rangle\langle a| e^{-iHL}|b\rangle \langle b| i \rangle=e^{-iE_\nu}\sum_a \left( \langle j| a\rangle e^{-i\frac{m_a^2}{2 E_\nu } L}\langle a| i\rangle \right);\notag \\
&&\langle j|a\rangle=(V_{PMNS})_{ja}
\label{oscillamp}
\end{eqnarray}

Specializing to two neutrinos, $\nu_{e,\mu}$, the appearence probability is given by:

\begin{equation}\begin{split}
&P(\nu_e \to \nu_\mu; E, L)= |A(\nu_e \to \nu_\mu)|^2=\cos^2\theta \sin^2\theta |1-e^{-i \frac{\Delta m^2 L}{2E_\nu}}|^2 \\
&=\sin^2(2\theta)\sin^2\left( \frac{\Delta m^2 L}{4E_\nu}\right),
\end{split}\end{equation}
where the energy is given in the ultrarelativistic limit
\begin{equation}
E\sim p+\frac{m^2}{2p}
\end{equation}

The flavor persistence probability is, of course:

\begin{equation}
P(\nu_e \rightarrow \nu_e ; E, L) = 1 -P(\nu_e \rightarrow \nu_{\mu} ; E, L).
\label{persistence}
\end{equation}

Numerically, the argument of the oscillating function reads:
\begin{equation}
\frac{\Delta m^2 L}{4E_\nu}=\frac{\Delta m^2 L}{4\hbar cE_\nu}\sim1.27\; \frac{\Delta m^2({\rm eV}^2) L({\rm km})}{E_\nu({\rm GeV})}=1.27\; \frac{\Delta m^2({\rm eV}^2) L({\rm m})}{E_\nu({\rm MeV})},
\end{equation}

For $\Delta m^2 \approx 1$~eV$^2$ and neutrinos of $1$~GeV, oscillations will take place in about $1$~km. 

Natural sources of varying distance to the detector have been used, like the Sun or the 1987 Supernova, for low-
energy neutrinos. High-energy neutrinos originate from the decay in flight of pions and muons produced by Cosmic Rays in the upper layer of the atmosphere (called atmospheric neutrinos). Artificial sources include nuclear reactors and high energy neutrino beams. 

A summary of sources and detection methods is given in Tabs.~\ref{low-energy-neutrinos} and~\ref{high-energy-neutrinos}.



\begin{table}[htb]\begin{tabular}{lccllll}\hline{\small  source }  &{\small production } &{\small 
$E_\nu$(MeV)} & {\small $L$(km) }  & {\small   reaction } &{\small detect.}&Exp. \\
& & & & at detector & meth. \\
\hline {\small nucl. reactor } & 
{\small $n\to \bar{ \nu}_e e^- p$}&{\small $1$}  & {\small $\sim 1$} & {\small $\bar{ \nu}_e p\to e^+ n$}& {\small scint. }  &Savannah \\
& & & & & & River\\ \\
{\small Sun (Be-B)}  & {\small  $\nu_e$ }&{\small $1-10$} & $1.4\cdot 10^{8} $& {\small $\nu_e \;{^{37}Cl}\to e\; {^{37}Ar}$}& 
{\small rc }&Homestake\\ \\
{\small Sun (p-p)}  & {\small $\nu_e$ }&{\small $0.2-0.7$} & $1.4\cdot 10^{8} $& {\small $\nu_e\; 
{^{71}Ga}\to e\; {^{71}Ge}$}& {\small rc  }&GALLEX,\\
& & & & & & SAGE\\ \\
{\small Sun (B)}  & {\small $\nu_e$ }&{\small $5.5-10$} & 
$1.4\cdot 10^{8} $& {\small $\nu_e\; p\to e\;n$}& {\small Ch.  }&Kamiokande \\ \\
{\small Sun (B)}  & {\small $\nu_e$ }&{\small 
$6-10$} & $1.4\cdot 10^{8} $& {\small $\nu \; d\to \nu\;p\;n$}& {\small Ch.  }&SNO \\ \\
{\small Supernova} & {\small $e\; p 
\to n\; \nu_e$} & {\small $1$} & {\small $1.7\cdot 10^{18}$} &{\small $\nu_e\;Nucl. \to e+\cdots$}&{\small  Ch. }&Kamiokande II \\ 
1987 & & & & & &\\ \\
nucl. reactor & $n\rightarrow \bar{\nu}_e e^{-}p$ & 1 & $\sim 1$& $\bar{\nu}_e p \rightarrow e^+ n $ & scint. & Chooz,\\ 
& & & & & & Daya Bay\\
\hline
\end{tabular}
\caption{\label{low-energy-neutrinos} {\rm Artificial and natural sources of low-energy neutrinos and methods to detect them; {\it scint.}, {\it Ch} and {\it rc} stand for scintillator or Cherenkov detector and radiochemical method, respectively. Table from Ref.~\cite{benhar1}. }}
\end{table}

\begin{table}[htb]\begin{tabular}{lllllll}\hline{\small source }  &{\small production } &{\small $E_\nu$(MeV)} & {\small $L$(km) }  & 
{\small   reaction} &{\small detect.}&Exp \\
& & & & at detector & meth. \\
\hline
{\small Atmosph. } & 
$\left(\begin{array}{c}
\pi \to \mu \nu_\mu \\ 
\mu \to \nu_\mu e \nu_e
\end{array}\right)$
& $10^3$ & {\small $\sim 20$}& {\small $\nu_{\mu/e} \; Nucl. \to \mu/e +\cdots$}&{\small Ch.}& Kamiokande\\
(zenith) & & & & &\\ \\
{\small Atmosph.}  &
$\left(\begin{array}{c}
\pi \to \mu \nu_\mu \\ 
\mu \to \nu_\mu e \nu_e
\end{array} \right)$
&{\small 
$10^3$} & {\small $\sim 13000$}& {\small $\nu_{\mu/e} \; Nucl. \to \mu/e +\cdots$}&{\small Ch.}& Kamiokande\\
(nadir) & & & & &\\ \\
{\small Acc.}  &{\small $\pi/K \to \mu \nu_\mu$}&{\small $10^{3-5}$}  & $0.1-1$ & {\small $\nu_\mu (\bar{ \nu}_\mu) 
\;Nucl. \to l^\mp+\cdots$ }& {\small imag. }&\\
(short base) & & & & &\\ \\
 {\small Acc.}  &{\small $\pi/K \to \mu \nu_\mu$} &{\small $10^{3-4}$}  & 
$300-900$ & {\small $\nu_\mu (\bar{ \nu}_\mu) \; Nucl. \to l^\mp+\cdots$} & {\small imag. }& JP, IT, USA\\
(long base) & & & & &\\
\hline
\end{tabular}
\caption{\label{high-energy-neutrinos} {\rm Artificial sources, natural sources and methods to detect high-energy neutrinos; {\it imag} stands for imaging detection methods. Table from Ref.~\cite{benhar1}.}}
\end{table}

\subsection{\bf Solar neutrinos}
Experiments to study neutrino oscillations have been performed since 1970. Tab \ref{Exp} gives the observed solar neutrino deficit,  namely the ratio of the observed flux from charged current processes to the estimated flux, using the Standard Solar Model.


\begin{figure}[!h]
        \centering
        {
       \includegraphics[scale=0.2]{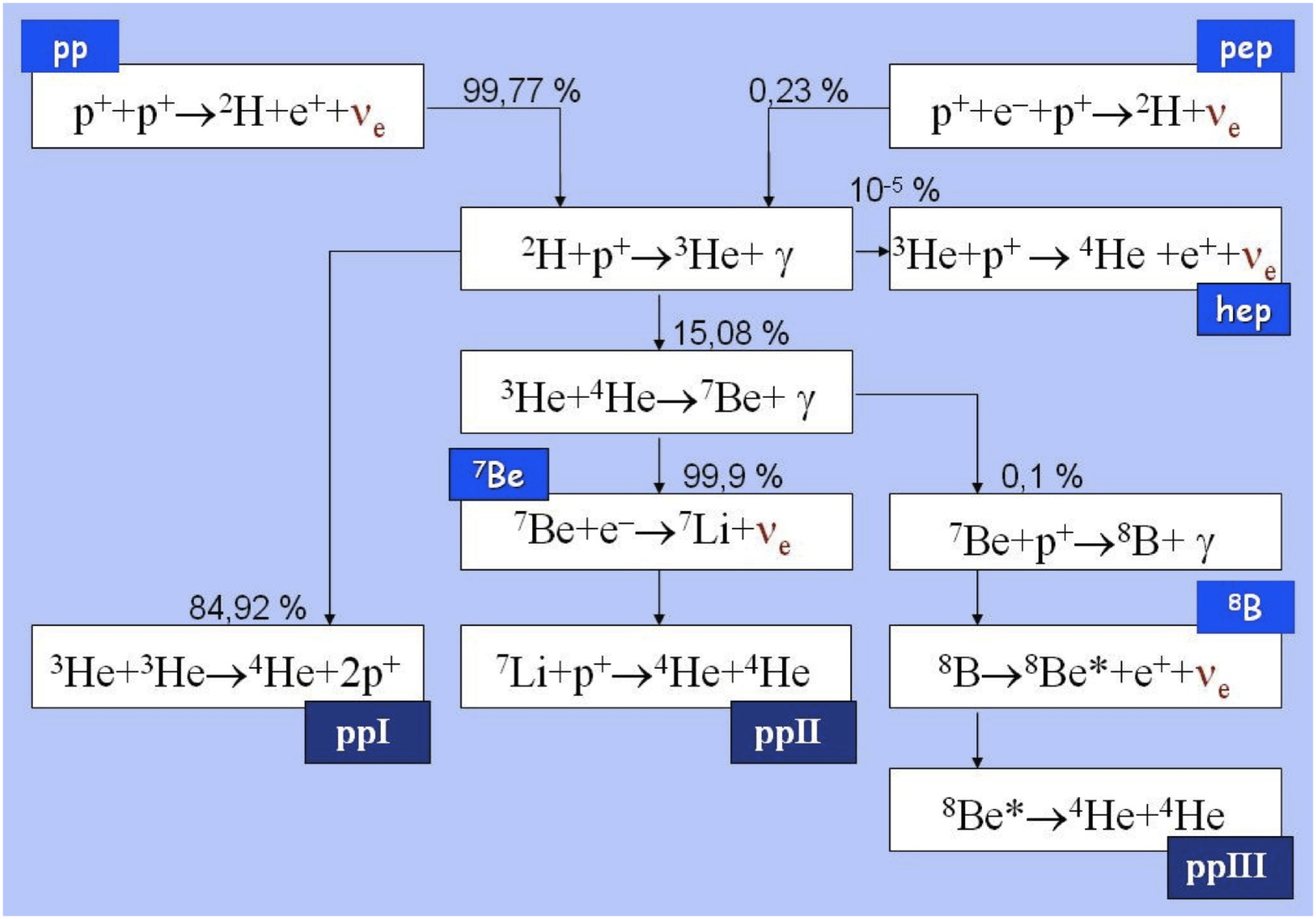}}
	   {
	     \includegraphics[scale=0.23]{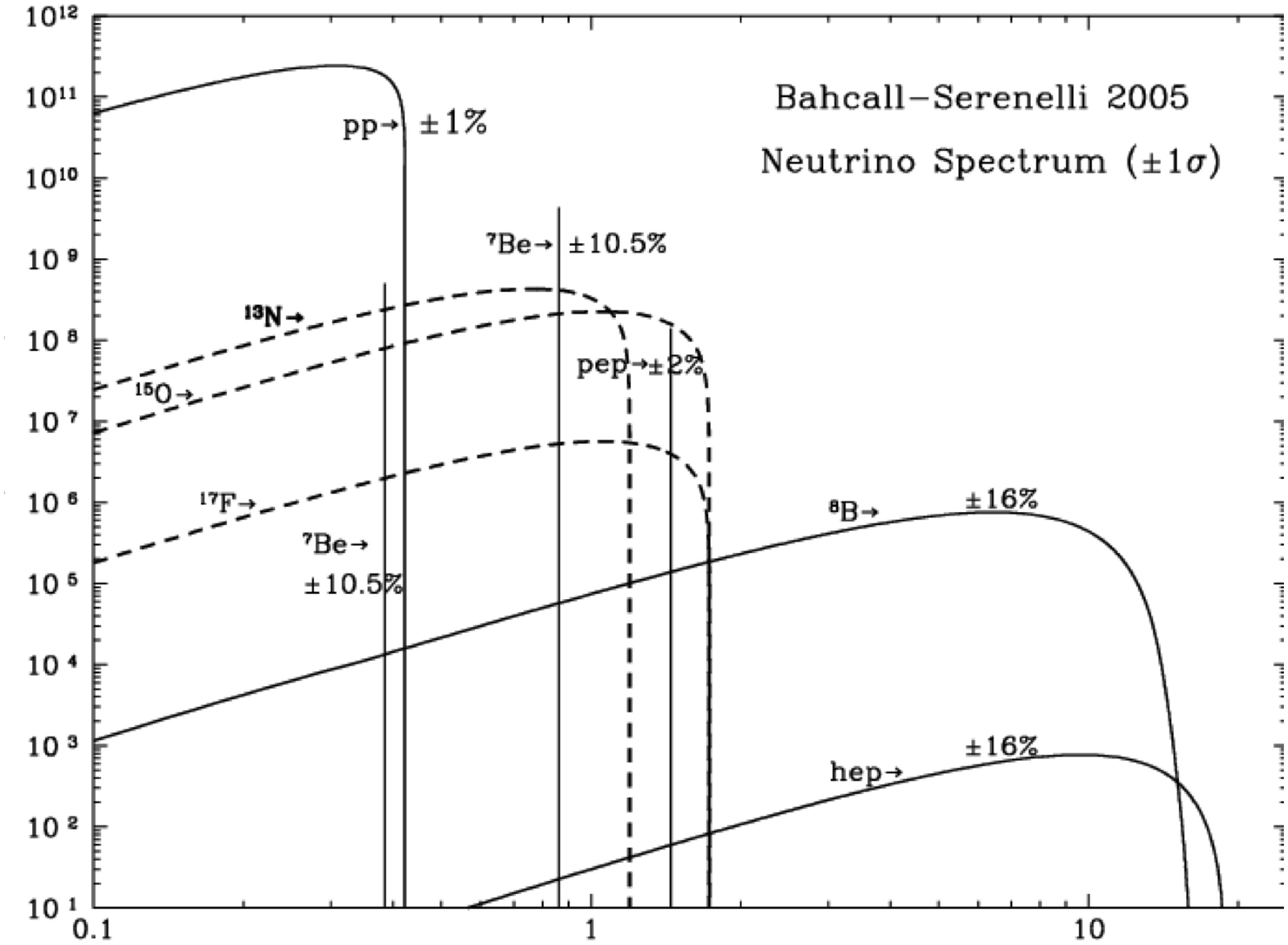}}
  	   \caption{ \label{solarspectrum} {\footnotesize (a) The Bethe cycles that produce the energy of the Sun by nuclear fusion of the light elements. (b) Energy spectra of neutrinos arising in the solar reactions. Figure from Ref.~\cite{Bahcall:2004pz}.
	   }}
\end{figure}
\begin{figure}[!h]
        \centering
       \includegraphics[scale=0.3]{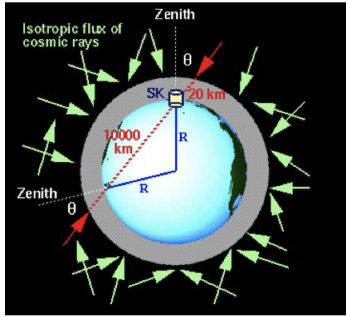}
  	   \caption{{\footnotesize S-K observation of atmospheric neutrinos with different path lengths. Figure: SuperKamiokande.}}
  	   \label{cosmicrays}
\end{figure}

\begin{table}[!h]
\begin{center}
\begin{tabular}{ccc}
Experiment &Observed/Expected & Years of observation \\
\hline
Homestake & $0.33 \pm 0.03 \pm 0.05$ & $1970-1995$ \\
Kamiokande & 0.54 $\pm 0.08^{+0.10}_{-0.07}$& $1986- 1995$\\
SAGE &  $0.50 \pm 0.06 \pm 0.03 $& $1990-2006$\\
GALLEX &$ 0.60 \pm 0.06 \pm 0.04$ & $1991-1996$\\
Super-Kamiokande & $0.456 \pm 0.005^{+0.016}_{-0.015}$ & $1996-$ \\
\hline
\end{tabular}
\end{center}
\caption{\label{Exp}{\rm  {\small Observed deficit in solar neutrinos experiments. For SNO, see Fig.~\ref{SNO}. Table from Ref.~\cite{benhar1}.}}}
\end{table}

Pontecorvo offered a radical, for the time, interpretation of the solar neutrino deficit in charged current processes. 

Solar neutrinos start as low-energy neutrinos and are initially  of pure $\nu_e$ flavor. Fusion reactions giving rise to neutrinos in the Sun have been identified by H. Bethe and are reported in Fig.~\ref{solarspectrum}, with the corresponding energy spectra. While traveling to the Earth, they undergo $\nu_e$-$\nu_{\mu}$ oscillations; but $1$ MeV muonic neutrinos do not have enough energy to produce the final muon in a charged current 
interaction, so detectors cannot see what Pontecorvo called sterile neutrinos
\footnote{the same applies if $\nu_e$ oscillates in a superposition of $\nu_\mu$ and $\nu_\tau$, as it happens with three lepton flavors. Another addition to Pontecorvo's simple picture is the possibility that neutrino oscillations are also induced by the interaction with the Sun's atmosphere~\cite{Wolfenstein:1977ue,Mikheyev:1989dy} considered below. }
 (i.e. unable to produce charged current reactions). Hence the measured flux of $\nu_e$ is  {\it less} than what would be expected from solar models. 

This interpretation could explain the deficit observed by R. Davis and Collaborators in  the Homestake experiment~\cite{homestake} and confirmed by successive experiments. The Gallex-GNO~\cite{gallex} and SAGE~\cite{sage} experiments are particularly significant in that they were sensitive to the low-energy neutrinos produced in the proton-proton (PP) cycle. PP neutrinos are by far the most abundantly produced neutrinos and their flux can be reliably determined from the energy produced by the Sun.

Later, the Subdury Neutrino Observatory (SNO) experiment was performed, which was sensitive to neutral current processes as well~\cite{Aharmim:2009gd}. The SNO results, shown in Fig.~\ref{SNO}, indicate that a solar deficit does not exist for neutral current processes, a clear footprint of oscillations~\cite{Nakamura:2010zzi}.

\paragraph{{\bf The MSW effect}} We can fit solar neutrinos with one mixing angle, $\theta_{12}$  and one mass-squared difference, $\Delta m^2_{12}$, since, {\it a posteriori}, we know that $\theta_{13} $, defined in (\ref{nu_e}), is much smaller than the other angles. However, for a quantitative analysis, we have to include the propagation of neutrinos through solar atmosphere, which may give rise to what is called the WMS resonant effect~\cite{Wolfenstein:1977ue,Mikheyev:1989dy}, followed by propagation {\it in vacuo} from Sun to Earth.

The effect of the solar atmosphere depends from the neutrino energy, as discussed e.g. in~\cite{Nakamura:2010zzi}, and this accounts for the different ratios found in Tab.~\ref{Exp}.
\begin{itemize} 
\item in the Gallium experiment (low energy, $p-p$, neutrinos) the MSW effect is negligible and the ratio in the table is simply the long distance average of eq.~(\ref{persistence}), namely
\begin{equation}
R(Ga)\approx0.60=1-\frac{1}{2}\sin^2(2\theta_{12})~\to \sin^2(\theta_{12}) \approx 0.27\nonumber
\end{equation}

\item in the Homestake experiment (high energy, $B-Be$, neutrinos) the MSW is such that the neutrino emerges from the solar atmosphere in the higher eigenstate, $\nu_2$, and then travels undisturbed to the Earth, so that:
\begin{equation}
R(B-Be)\approx 0.33=\sin^2(\theta_{12}) \nonumber
\end{equation}
\end{itemize}

It is remarkable that the different ratios for $p-p$ and for $B-Be$ neutrinos are reproduced by approximately the same mixing angle. 


\subsection{\bf Reactor antineutrinos} 
\begin{figure}[!h]
        \centering
      {
        \includegraphics[scale=0.45]{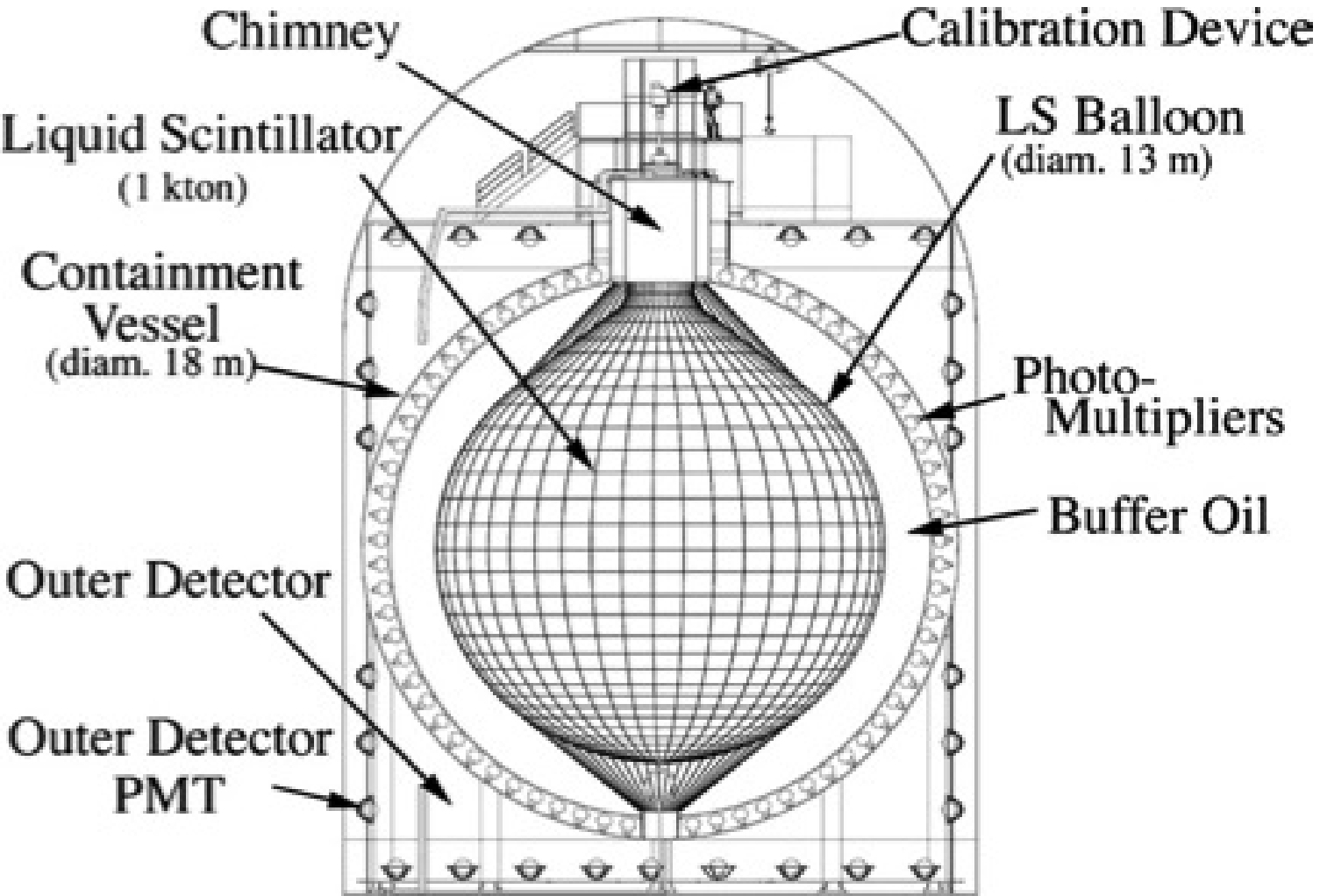}} 
        {
        \includegraphics[scale=0.3]{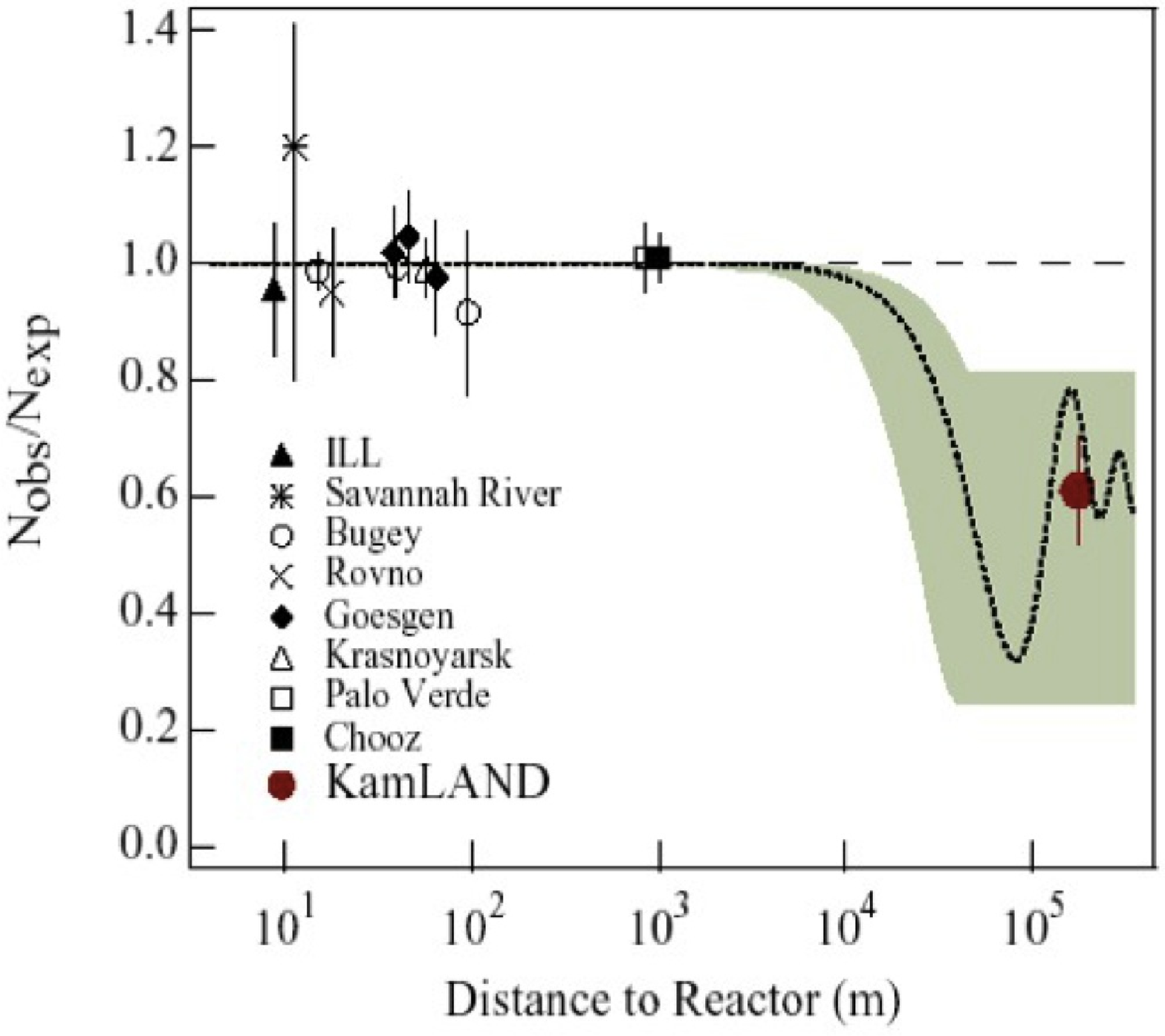}}
  	   \caption{(a) Schematic diagram of the KamLAND detector. (b) The ratio of measured to expected antineutrinos flux from reactor experiment. The solid circle is the KamLAND result plotted at a flux-weighted average distance of about 180$km$. The shaded region indicates the range of flux predictions corresponding to the 95$\%$C.L. large mixing angle region (LMA) from a  global analysis of the solar neutrino data. The dotted curve 
	   is representative of the best-fit LMA prediction and the dashed curve is expected for no oscillations. Figure from Ref.~\cite{Eguchi:2002dm}.}
  	   \label{KamLandDataplot}
\end{figure}

\begin{figure}[!h]
        \centering
         {
       \includegraphics[scale=0.35]{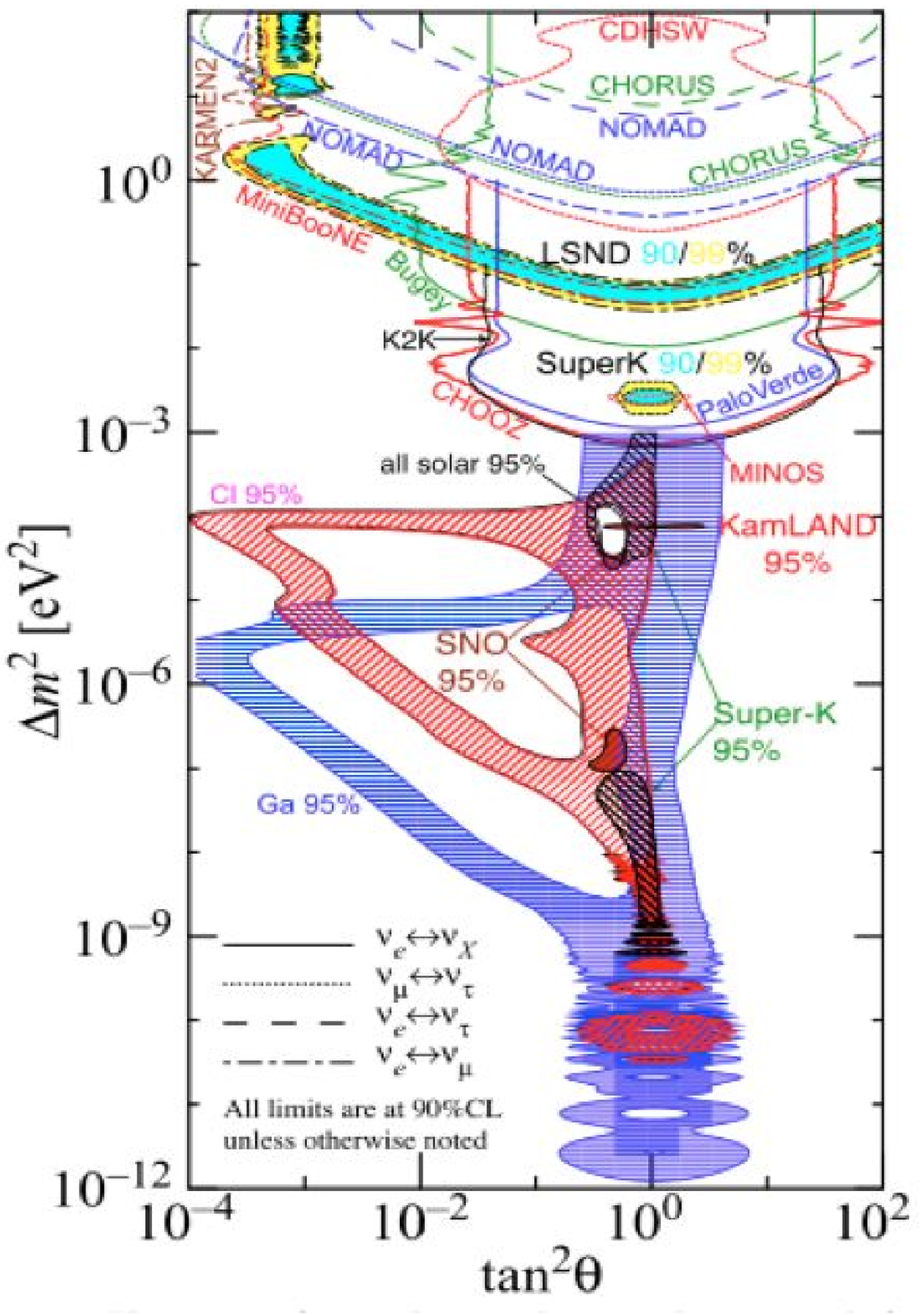}}
	    {
	     \includegraphics[scale=0.30]{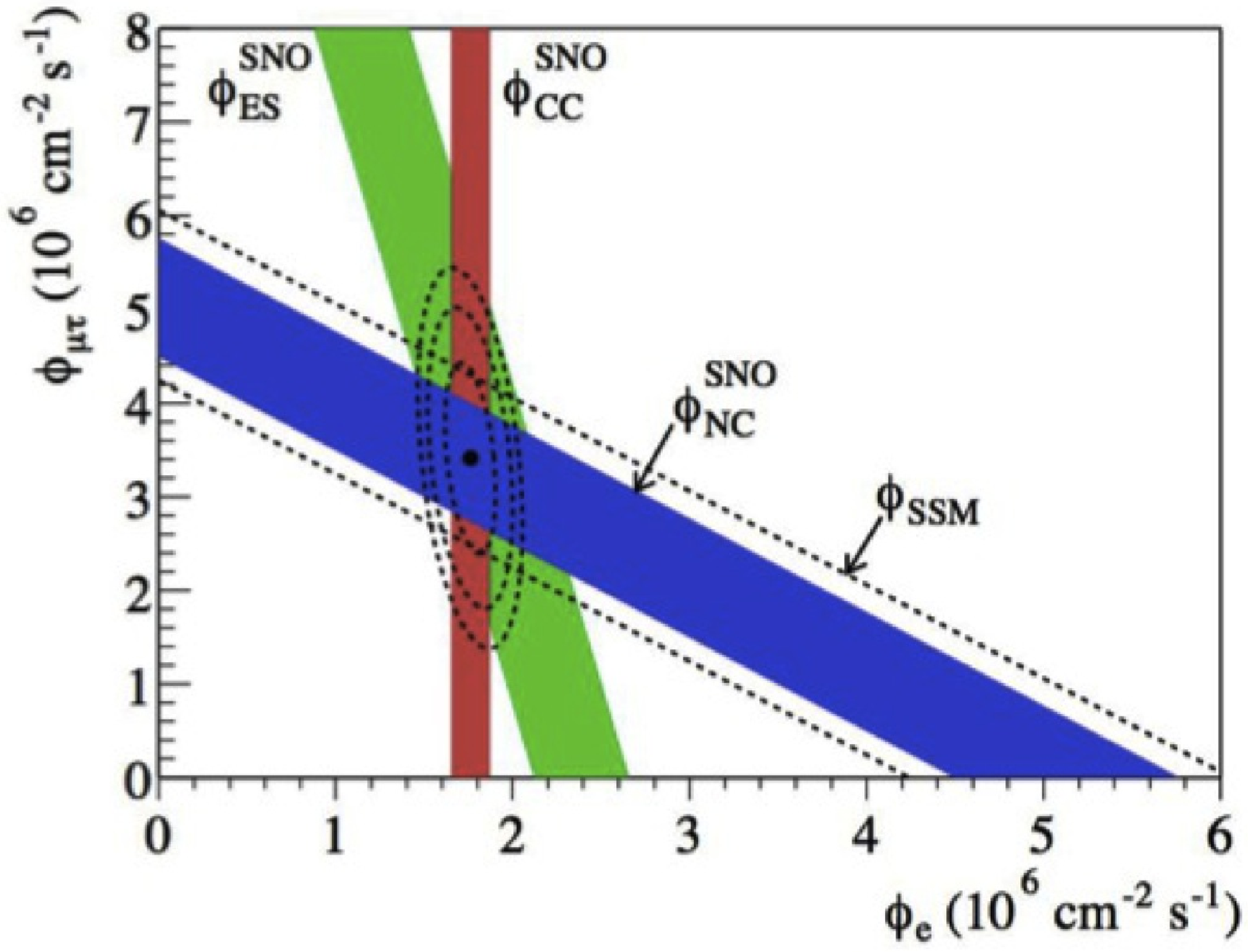}}
  	   \caption{\label{SNO} {\small (a) Fit to the currently available data for electron neutrino oscillations. The lines on the upper part indicate excluded regions from earlier experiments, the filled regions are allowed values. KamLAND 95$\%$ C.L. area is red, and SNO is brown. KamLAND chooses definite values for $\sin^2 \theta$ and $\Delta m^2$ for solar neutrinos. Figure from Ref.~\cite{Nakamura:2010zzi}. 	   
	   (b) Neutrino fluxes from SNO. The x-axis shows electron-neutrino flux, the y-axis flux of other neutrinos (it is not possible to distinguish $\mu$ and $\tau$). The red band shows the result of the charged-current analysis (CC), sensitive to electron-neutrinos only. The blue band is the neutral-current (NC) analysis, equally sensitive to all types. The green band is elastic $\nu_e$ scattering (ES), which prefers electron-neutrinos but has some sensitivity to other types. The dashed line is the total neutrino flux expected in the Standard Solar Model (SSM). Figure from Ref.~\cite{Aharmim:2009gd}.} 
	   }
	   \end{figure}

Oscillations of reactor antineutrinos have been observed by the experiment KamLAND (Kamioka Liquid Scintillator Antineutrino Detector). 
KamLAND is an experimental device (see Fig. \ref{KamLandDataplot} (a)) that was built at the Kamioka Observatory, an underground 
Neutrino Observatory near Toyama, Japan, see Ref.~\cite{Eguchi:2002dm}. It receives neutrinos from the $55$ Japanese nuclear power reactors, which are isotropic $\bar{\nu}_e$ 
sources.  KAMLAND observes a flux of antineutrinos which is definitely below the level observed in experiments close to the reactors. Fitting the flux of antineutrinos to the known spectra and distances of individual reactors, the experiment obtained the very remarkable oscillation signal shown in Fig.~\ref{KamLandDataplot}(b), corresponding to the parameters~\cite{Abe:2008aa}:
:
\begin{equation}
\tan^2 \theta_{12} = 0.47^{+0.06}_{-0.05},~\Delta m^2_{12} = 7.59\pm 0.21 \times 10^{-5}~ {\rm eV}^2.
\end{equation}
which falls inside the region of the parameters allowed by solar neutrino data, Fig.~\ref{SNO}-a, and it allows to choose a definite solution with a small error.

\subsection{\bf Atmospheric neutrinos} The underground installation Super-Kamiokande (S-K) detects neutrinos originated from the decay of pions and muons produced in the atmosphere by the interactions of high energy. These neutrinos go through the Earth without attenuation, so S-K can compare neutrinos produced at the zenith and arriving directly to SK, with neutrinos coming from below, which have been produced at the other side of the Earth and have traveled without appreciable attenuation over distances of the order of $10.000$~km, see Fig. \ref{cosmicrays}. 

Surprisingly, muon neutrinos coming from below are reduced with respect to those coming from above (about $50 \%$ less). In 1997, the disappearence of muon neutrinos coming from the other side of the Earth, was definitely confirmed. The phenomenon has been interpreted as the oscillation of muon neutrinos into $\tau$~neutrinos, and it gave the first experimental evidence of oscillations of terrestrial neutrinos. 

\subsection{\bf The OPERA $\tau$ neutrino events}
Atmospheric muon neutrinos do not have enough energy to produce $\tau$ leptons in charged current processes. To observe directly the transformation $\nu_\mu\to \nu_\tau$ CERN has built a beam of essentially  muon neutrinos, directed towards the Gran Sasso Laboratory, with energy above the threshold for producing $\tau$ leptons in charged current interactions and to study directly the oscillation $\nu_\mu \to \nu_\tau$. 


The OPERA collaboration at the Laboratori Nazionali del GranSasso has built and operated a large scale hybrid detector made by iron plates, where neutrino interactions may take place, separated by layers of special photographic emulsions, where the traces of the particles produced can be visualized with resolution high enough as to make possible to separate the $\tau$ decay vertex from the neutrino interaction point~\cite{Acquafredda:2009zz}. 

The chain of events that have been detected is:
\begin{eqnarray}
&& \nu + {\rm Nucleus} \to \tau + {\rm anything}\notag \\
&&\tau \to \nu_\tau + {\rm detected~particles}
\end{eqnarray}
the second event taking place at a detectable distance from the first, primary, event.

OPERA has observed, until now, four events of charged current neutrino interactions with a visible decay of a $\tau$ lepton in the final state~\cite{Agafonova:2010dc}. These events are above the possible background constituted by charm production with subsequent charm semileptonic decay with a low energy muon~\cite{Dzhatdoev:2014qsa}, providing an evidence at $4.2~\sigma$  confidence level~\cite{delellis} that neutrinos born in CERN as muon neutrinos transform in $\tau$ neutrinos after their $730$~km journey from CERN to LNGS. 

\begin{figure}[h]
        \centering
        {
       \includegraphics[scale=0.2]{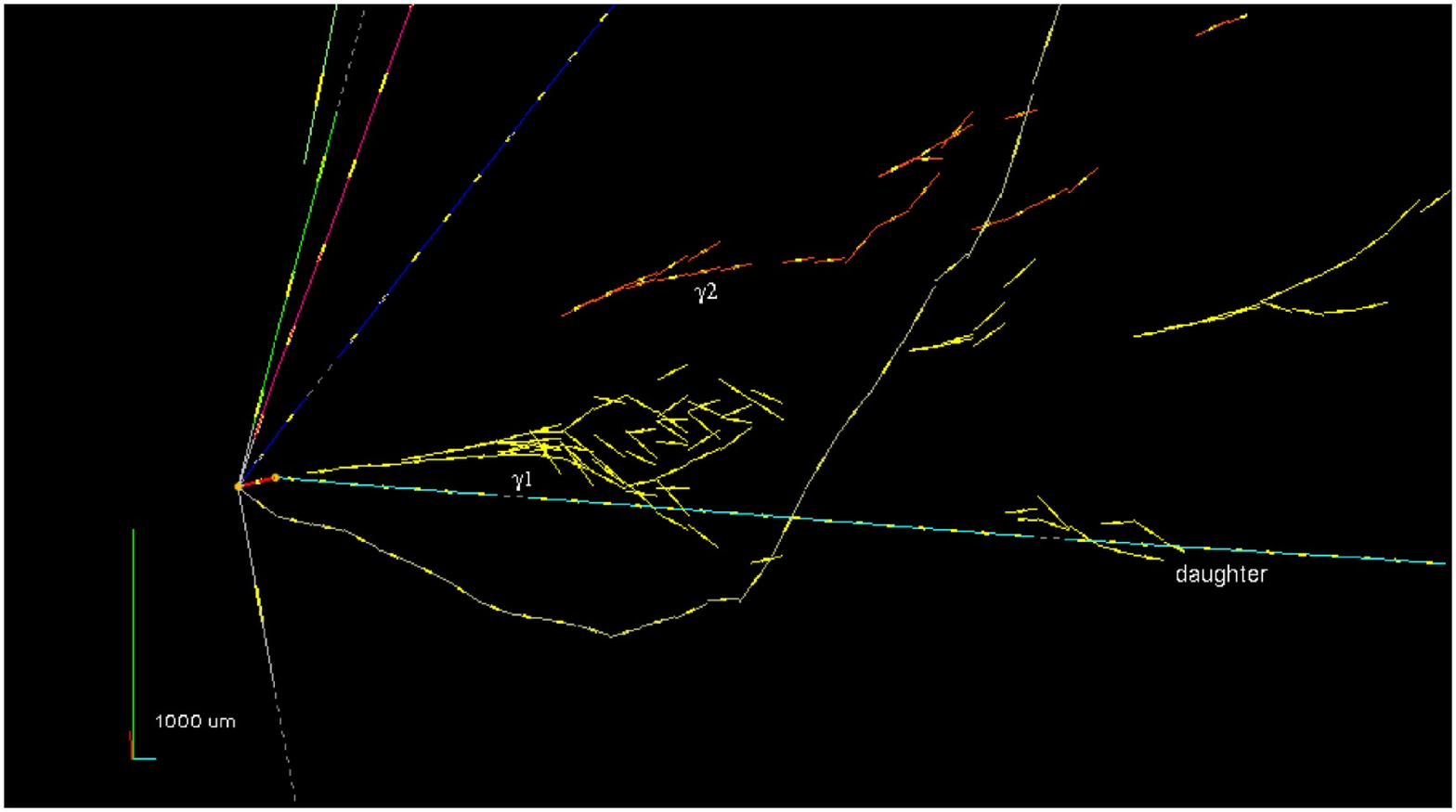}}
       {
        \includegraphics[scale=0.18]{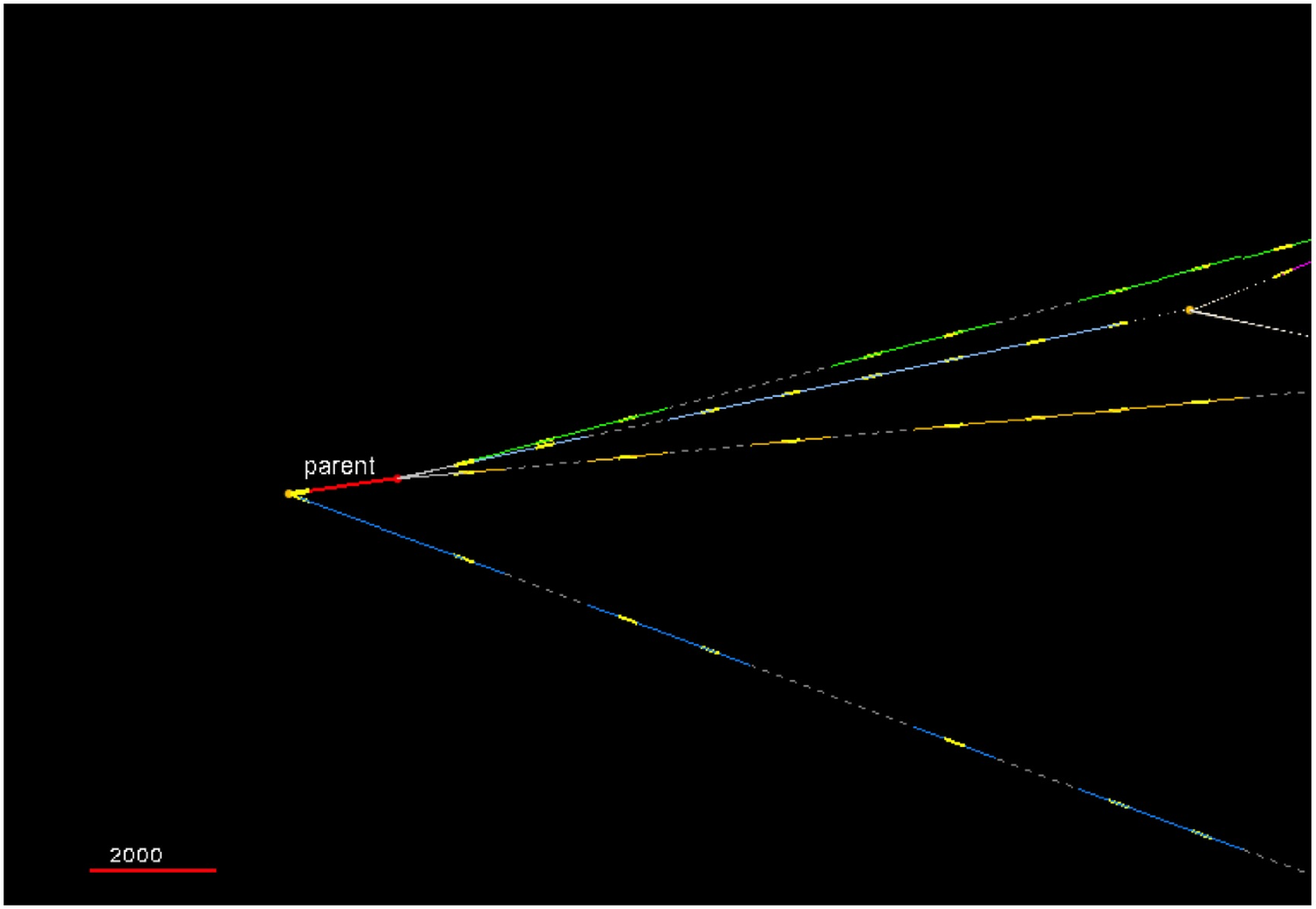}}
  	   \caption{\label{tau-decay12} Opera events 1 and 2. Primary and decay  vertices are joined by a red line: (a) $ \tau^{-} \rightarrow \rho^{-} \nu_{\tau}$~followed by $ \rho^{-}  \rightarrow \pi^{0} \pi^{-}$; (b) $\tau \rightarrow \nu_{\tau} + \nu_{\mu} + \mu$. Figures from Ref.~\cite{Agafonova:2010dc}.}
      {
       \includegraphics[scale=0.18]{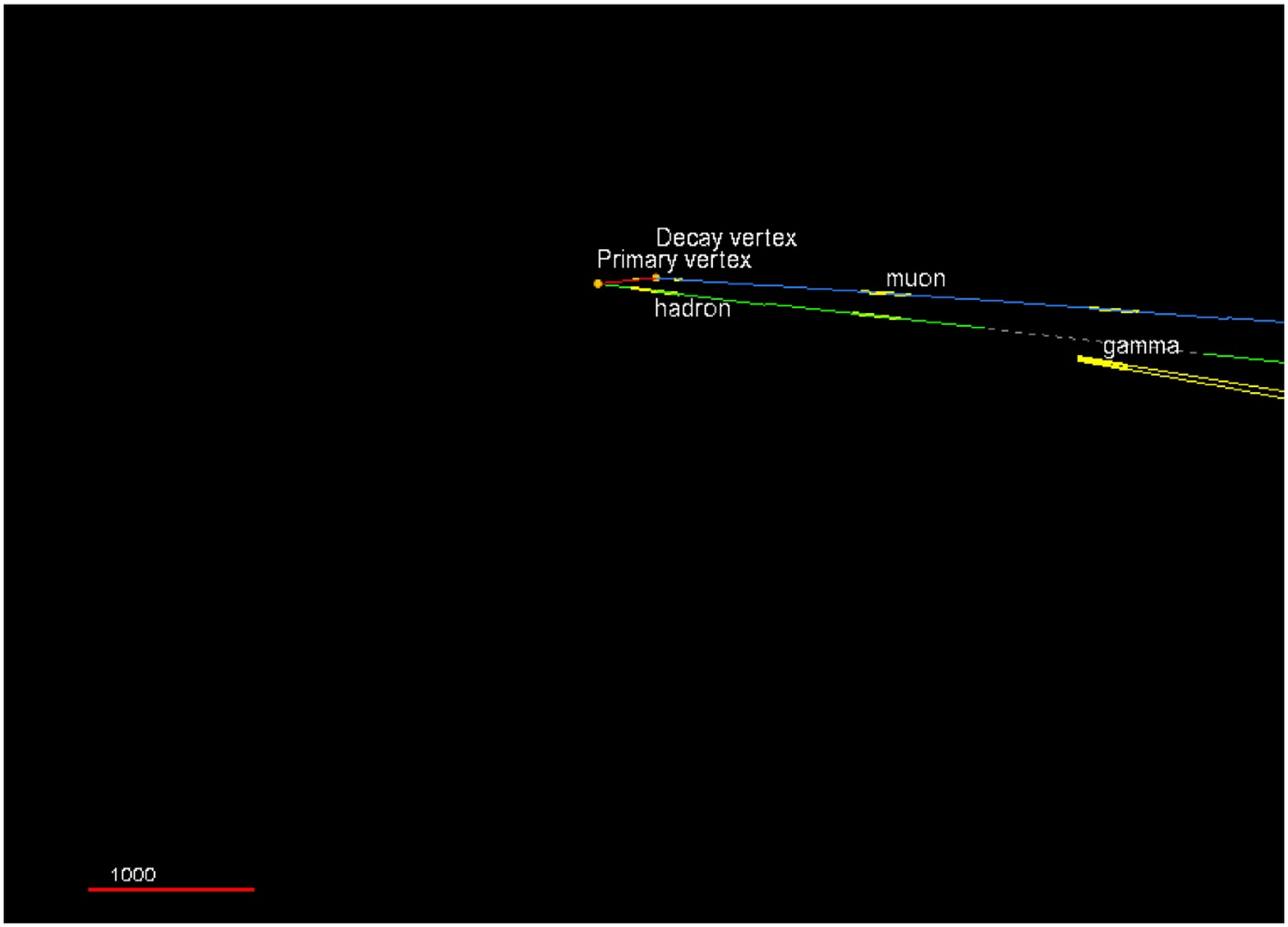}}
       {
        \includegraphics[scale=0.18]{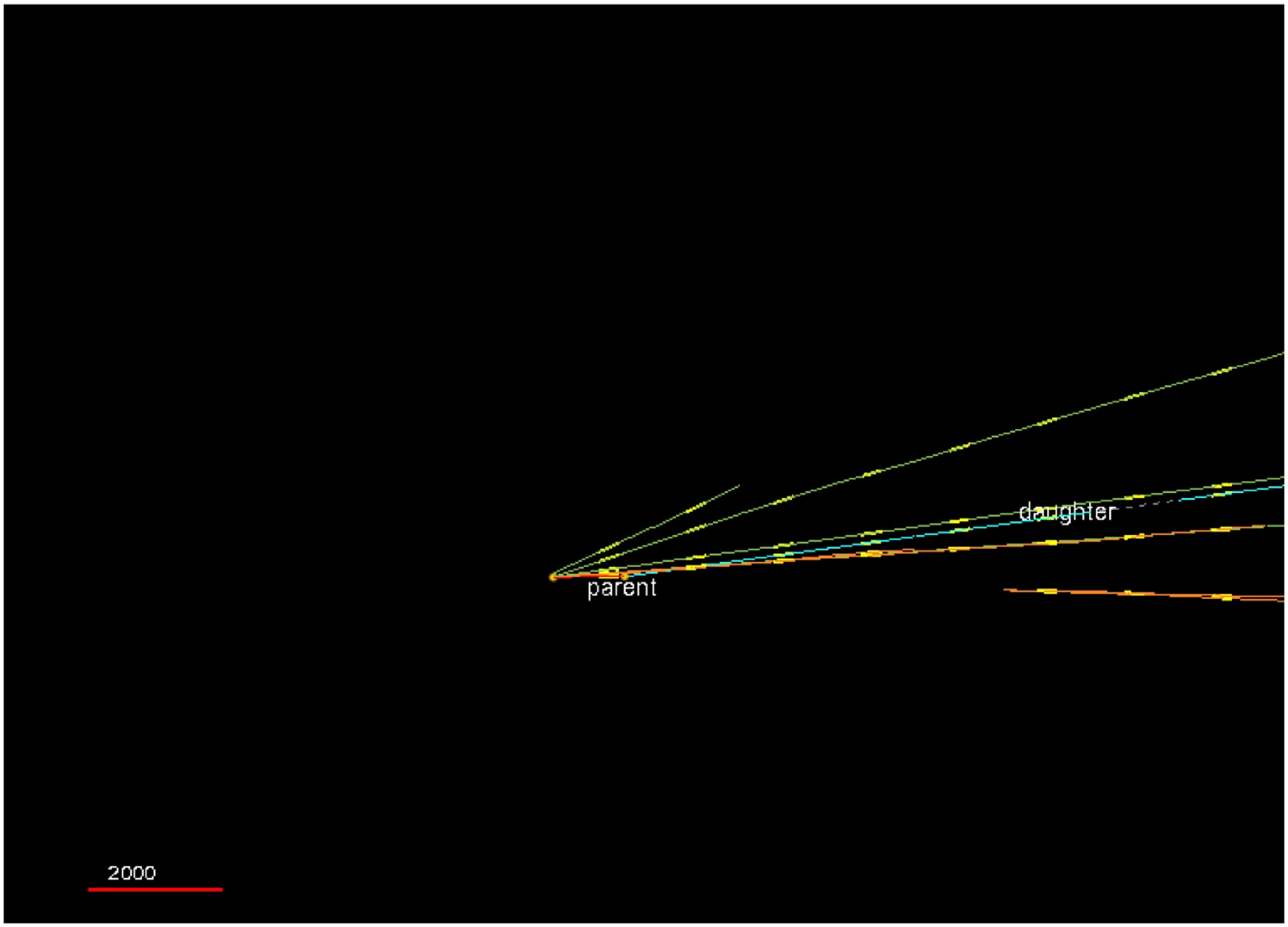}}
  	   \caption{\label{tau-decay34} Opera events 3 and 4: (a) $\tau \to \nu_{\tau} + 3 \; {\rm hadrons}$; (b) $\tau \to \nu_{\tau} + 1~ {\rm pion}$. Figures from Ref.~\cite{Agafonova:2010dc}.}
\end{figure}
A brief description of the events is as follows.
\begin{itemize}
\item Fig~\ref{tau-decay12}(a):
\begin{equation}
 \begin{split}
& \tau^{-} \rightarrow \rho^{-} \nu_{\tau},\notag \\
& \rho^{-}  \rightarrow \pi^{0} \pi^{-},\notag \\
& \pi^{0} \rightarrow  \gamma \gamma.\notag
\end{split} 
\end{equation}
It is possible to distinguish the primary vertex and the decay vertex. Two gamma rays point to the secondary vertex, signalling the $\pi^{0}\to \gamma \gamma$ decay, the line labelled with ``daughter'' is the $\pi^{-}$.

\item Fig.\ref{tau-decay12}(b), represents a $\tau$ muonic decay:
\begin{equation}
\tau \rightarrow \nu_{\tau} + \nu_{\mu} + \mu,
\end{equation}

and again it is possible to see the primary and secondary vertex, where an energetic muon comes out. 

\item  Fig.~\ref{tau-decay34}(a), features a secondary vertex with three particles which are interpreted as hadrons:  
\begin{equation}
\tau \rightarrow \nu_{\tau} + 3 \; {\rm hadrons},\notag
\end{equation}

\item  Fig.~\ref{tau-decay34}(b), is interpreted as:
 \begin{equation}
\tau \rightarrow \nu_{\tau} + \nu_{\mu} +\pi \notag
\end{equation}
\end{itemize}

\subsection{\bf The last real angle, $\theta_{13}$}

The latest experiment for neutrino oscillations is the Daya Bay (China) reactor experiment~\cite{dbchina}, which allows to measure the angle $\theta_{13}$.
It is an experiment with near detectors close to the reactor units and far detectors located at $L=1.6$~km from the nuclear reactors. 

It is instructive to see how the result may produce a value for $\theta_{13}$.

We consider three neutrinos and neglect all angles axcept $\theta_{13}$ and $\theta_{23}$. In this approximation:
\begin{equation}
\Delta_{13}= - \Delta_{31} \approx \Delta_{23} = - \Delta_{32} 
\end{equation}
Under these conditions it is easy to compute the probability $P(\nu_e \rightarrow \nu_{e}, L)$:
\begin{equation}\begin{split}
P(\nu_e \rightarrow \nu_{e}, L)&=\sum_{a,b}~e^{i\Delta_{ab}} |U_{ea}|^2 ~|U_{eb}|^2=\notag \\
&=\sum_a |U_{ea}|^4 + 2 |U_{e1}|^2 |U_{e2}|^2+(e^{i\Delta_{13}}+e^{-i\Delta_{13}})\left[|U_{e1}|^2 |U_{e3}|^2+|U_{e2}|^2 U_{e3}|^2\right]=\notag\\
&=c_{13}^4+s_{13}^4+s_{13}^2 c_{13}^2\cdot 2 \cos\Delta_{13},
\end{split}
\end{equation}
and then
\begin{equation}\label{Ptheta_13}
P(\nu_e \to \nu_e,L)=1-\sin^2 2\theta_{13} \sin^2\frac{\Delta m^2_{23}L}{4E_\nu},
\end{equation}

\begin{figure}[!h]
        \centering
       \includegraphics[scale=0.5]{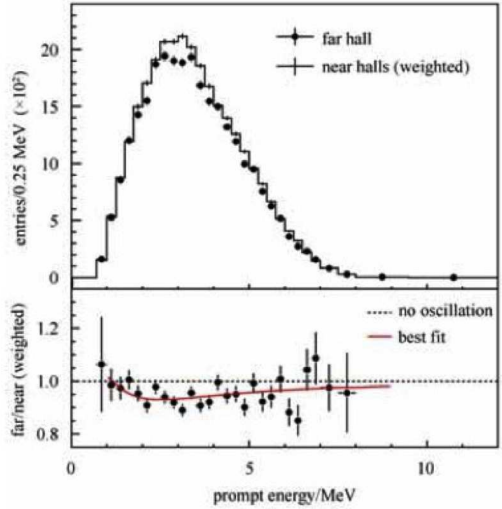}
  	   \caption{Neutrino spectrum from Daya Bay experiment. Figure from Ref.~~\cite{dbchina}.}
  	   \label{DayaBay}
\end{figure}

The simple expression (\ref{Ptheta_13}) allows to determine $\theta_{13}$ since we know all the other parameters. 
Fig.~\ref{DayaBay} shows the Daya Bay data for the neutrino's energy spectrum, obtained by comparing the fluxes at the near and at the far detectors. The oscillation reflects in an energy modulation that determines the angle $\theta_{13}$.

\subsection{\bf Summing up}
A global fit to neutrino masses and oscillations has been recently presented in Ref.~\cite{Capozzi:2013csa}.

There are two mass differences with three neutrinos, data are compatible with the two mass-squared differences ($1\sigma$ errors): 
\begin{equation}
| \Delta m_{12}^2|=(7.54 \pm 0.24) \cdot 10^{-5} eV^{2}  \hspace{5mm},~ |\Delta m_{23}^2|=(2.43 \pm 0.06) \cdot 10^{-3} eV^{2},
\end{equation}

The angles for solar neutrinos, $\theta_{12}$ ($\nu_{e} \rightarrow \nu_{\mu,\tau}$), and atmospheric neutrinos, $\theta_{23}$ ($\nu_\mu\rightarrow \nu_\tau$), are: 
\begin{equation}
 \sin^2\theta_{12}=0.308\pm 0.17\hspace{5mm},~ \sin^2\theta_{23}=0.437\pm0.28.
\end{equation}
The last angle, $\theta_{13}$ determined by the Daya Bay collaboration, is: 
\begin{equation}
sin^2 \theta_{13}=0.023 \pm 0.004,
\end{equation}
The above values produce the numerical PMNS matrix anticipated in (\ref{pmatrix}).

\section{{ See-saw Neutrinos in Three Generations}}
\label{seesawnu}

We have already discussed the see-saw mechanism in the case of one neutrino generation. Now we turn to the case of three generations and the corresponding flavor symmetry. 

In general, we consider the Lagrangian for Yukawa interaction 
\begin{equation}
{\cal L}_Y={\cal L}_{quark}+{\cal L}_{ch.lept}+{\cal L}_{nu},
\end{equation}
${\cal L}_{quark}$ is the Lagrangian we have considered in Sect.~\ref{qm&m}, Eq.~(\ref{ylag}). The charged lepton Lagrangian is, analogously:
\begin{equation}
{\cal L}_{ch.lept}=[{\bar \ell}_L {\it Y}_E H E_R+{\rm h.c.}],
\end{equation}
where $\ell_L$ is the left-handed doublet
\begin{equation} \ell_L = 
\begin{pmatrix}
\nu_L \\
E_L
\end{pmatrix},
\end{equation}
Now $\ell_L$  represent the three left-handed neutrino and charged lepton left-handed generations and $Y_E$ determines the interaction of $\bar{\ell}_L$ with the Higgs boson H and the charged right-handed leptons $E_R$. 

${\cal L}_{nu}$ is neutrinos Lagrangian, which we write as:
\begin{equation} \label{L nu}
{\cal L}_{\nu}=\frac{M}{2}N\gamma_0 N+[{\bar \ell}_L {\it Y}_\nu {\tilde H}  N+ {\rm h.c.}],
\end{equation}
Following Sect.~\ref{mechanismss} we have introduced three generations of heavy Majorana neutrinos, $N$, coupled to the left-handed leptons by the Yukawa coupling $Y_\nu$. $M$ is the Majorana mass of the heavy neutrinos, assumed to be degenerate.

The quark flavor group broken by ${\cal L}_{quark}$ was characterized in (\ref{gflav}) as  ${\cal G}_{quark} = SU(3)_q \otimes SU(3)_U \otimes SU(3)_D$.

For the leptons, we assume  

\begin{equation}
G_{lept} = SU(3)_\ell\otimes SU(3)_E\otimes O(3)_N
\label{gylept} 
\end{equation}
$SU(3)_\ell$ refers to the three generation doublets, $SU(3)_E$ to the right-handed, charged lepton, fields and $O(3)_N$ to the Majorana heavy neutrinos. 

For the first time, we see a difference between quark and lepton flavour group, implied by the see-saw mechanism and by the Majorana nature of $N$.

If we write $N= \nu_R + \nu_{R}^{\star}$, we obtain
\begin{equation} \label{L nu 2}
{\cal L}_{nu}=\frac{M}{2}\nu_R\gamma^0 \nu_R + {\bar {\ell}}{\tilde H}Y_\nu \gamma^0 \nu_R + {\rm h.c.}.
\end{equation}
Lagrangian (\ref{L nu 2}) has a quadratic plus linear term in $\nu_R$ and we obtain the effective low energy Lagrangian by performing a functional integral over $N$. To this aim, we shift $\nu_R$ by a field $A$ 
\begin{equation} \label{shift nu_R}
\nu_R \rightarrow \nu_R + A,
\end{equation}
and we choose $A$ so as to cancel the linear term in $\nu_R$ in (\ref{L nu 2}) and remain with a purely quadratic Lagrangian\footnote{recall that in the Majorana representation $\nu_R$ are anticommuting quantities, so that $\nu_R\gamma^0 \nu_R \neq 0 $.} 
\begin{equation}
{\cal L}_{nu} = \frac{M}{2}\nu_R\gamma^0 \nu_R +( M~A +{\bar {\ell}}{\tilde H}Y_\nu )\gamma^0 \nu_R + {\ell}^C_L\gamma^0 {\tilde H}Y_\nu A+  {\rm h.c.}.
\end{equation}
that is:
\begin{equation}
A=-\frac{1}{M}{\bar {\ell}}{\tilde H}Y_\nu
\label{shift} 
\end{equation}
After the shift~(\ref{shift nu_R}), the functional integral on $\nu_R$ becomes gaussian and  can be dropped. We remain with the effective see-saw Lagrangian:
\begin{equation} \begin{split}
 {\cal L}_{nu,~l.e.} &={\ell}^C_L\gamma^0 {\tilde H}Y_\nu A+ {\rm h.c.}.= \\
&=-{\ell}^C_L\gamma^0 {\tilde H}Y_\nu \frac{1}{M}Y_\nu^T {\tilde H}^T({\ell}^C_L)^T+ {\rm h.c.},
\end{split} \end{equation}
which contains the two doublets ${\ell}^C$, the two Higgs fields ${\tilde H}$ and the factor $\frac{1}{M}$, in accordance with the see-saw mechanism. In fact when we replace Higgs field with its vacuum expectation value we get the mass Lagrangian
\begin{equation}
{\cal L}_{mass}=\nu_L\gamma^0 M_\nu \nu_L + {\rm h.c.},
\end{equation} 
with the Majorana mass $M_\nu$ given by
\begin{equation}
M_\nu= \frac{v^2}{M}Y_\nu Y_\nu^T, \label{Majorana mass}
\end{equation}
$v^2$ is the square of the vacuum expectation value of the Higgs field and $Y_\nu$ is the Yukawa coupling. 

As in the quark case, by using the lepton flavor symmetry we can reduce the leptonic Yukawa coupling to a standard diagonal form. Diagonalization of $Y_E$ is obtained by bi-unitary transformations belonging to the flavor group:
\begin{equation}
Y_E \to U_{\ell} y_E U_{E}, 
\label{yEtransf}
\end{equation}
with $y_{E}$ diagonal, real and positive. As in the $up$ quark case, the matrices $U_{\ell}$ and $U_E$ can be reabsorbed into a gauge invariant field redefinition and disappear completely: we may take directly:
\begin{equation}
Y_E= y_E={\rm diag.} 
\label{yEtransf2}
\end{equation}

Group transformations on $Y_\nu$ are of the form:
\begin{equation}
Y_\nu \to U_{\ell} Y_\nu {\cal O}^T
\label{ynutransf}
\end{equation}
however, to obtain a complete diagonalization we have to perform a bi-unitary transformation,  so that
\begin{equation}
Y_\nu=U_L y_\nu  \omega  U_R, 
\label{ynudiag}
\end{equation}
$y_{E, \nu}$ being diagonal, real and positive matrices, $U_R$ and $U_L$ unitary matrices and $\omega$ a diagonal phase matrix of unit determinant (Majorana-phase matrix), which is essential for $y_\nu$ to be real, positive and diagonal. If we substitute (\ref{ynutransf}) in (\ref{Majorana mass}) the low-energy neutrino mass will be represented by the complex matrix:
\begin{equation}
M_\nu=\frac{v^2}{M}U_L(y_\nu \omega U_R U_R^T \omega y_\nu)U_L^T, 
\end{equation}

Being symmetric, $M_\nu$ is diagonalized according to\footnote{if $M_\nu =U\cdot$diag$\cdot V$ and $M_\nu^T=V^T\cdot$diag$\cdot U^T=M_\nu$, then $V=U^T$.}
\begin{equation}
M_\nu=U_{\rm PMNS} \; \Omega \; {\hat m}_\nu \; \Omega  \; U_{\rm PMNS}^T, 
\end{equation}
with the unitary, Pontecorvo-Maki-Nakagawa-Sakata, matrix $U_{PMNS}$. The diagonal Majorana-phase matrix $\Omega$ makes ${\hat m}_\nu$ a diagonal, real and positive matrix. 
Note that we consider the $Y$s as the fundamental variables and $Y Y^T$ as a derived quantity.

To count the real parameters appearing in the lepton Yukawa couplings eqs.~(\ref{ynudiag}) we start from neutrinos. We have $4$ parameters in $U_L$, as in the CKM matrix, $3$ real eigenvalues in $y_\nu$ and $3$ parameters in $U_R$ counted as follows: $8$ for a general $3\times 3$ special, unitary matrix, less $3$, corresponding to an orthogonal transformation we may perform on the Majorana fields, less $2$ phases we include in $\omega$. Adding the $3$ real eigenvalues of $y_E$, eq.~(\ref{yEtransf2}), we obtain a total of $15$ parameters. Coorespondingly,  we shall need as many invariants, see also~Ref.~\cite{Jenkins}. 

Note that the low-energy observable $M_\nu$, eq.~(\ref{masses}), contains $9$ parameters only ($4$ for the $U_{\rm PMNS}$ matrix, $3$ mass eigenvalues and $2$ Majorana phases). This is because we can factorize from ${\it Y}_\nu$ a complex orthogonal hermitian matrix, hence $3$ parameters,  which would drop from the 
expression in Eq.~(\ref{masses}),  see Ref.~\cite{Casas:2001sr}.

\vskip0.5cm

 \section{{ Yukawa Couplings as Fields}}
\label{dynyuk}

Until now we have considered the Standard Theory, where Yukawa couplings $Y$s are considered as fundamental constants. However, 
the universality of Yukawa couplings postulated by the Minimal Flavor Violation principle is difficult to reconcile with the idea that the $Y$s are 
``just'' renormalized constants. 

Froggat and  Nielsen~\cite{Froggatt:1998tj} introduced the idea the Yukawa couplings to be the vacuum expectation values of some new fields, which break spontaneously the flavor symmetry. 

Note that this is exactly what happens for quarks, whose mass depends on the value of the Higgs field. Now we repeat the same argument at a more fundamental level by saying that Yukawa couplings can be determined by a variational principle, i.e. by the minimum of a new ``hidden 
potential'' which is invariant under the flavor symmetry. 

This idea was considered in the late sixties by N. Cabibbo, as a possible way to explain the origin of the weak, Cabibbo, angle and the symmetry of the unknown potential was $SU(3)$ or chiral $SU(3)\otimes SU(3)$. The concept was explored by L. Michel and L. Radicati~\cite{Michel:1970mua}  in a more general group theoretical setting, and by Cabibbo and myself~\cite{Cabibbo:1970rza}.

If we follow this idea, we discover that there are certain minima of an invariant potential which are more {\it {\b natural}}. 

For  an example, take the case of rotational invariance. The potential must be a function of $r^2$ and therefore the  derivatives of the potential in $r=0$ are always equal to zero: there is always a natural extremum in $r=0$, determined by the 
rotation invariance of the potential. 

The point which was raised in the sixties was to ask if a potential invariant under a more complicated group, i.e. $SU(3)$, could have minima (or extrema) which are more natural than others and perhaps the ones chosen by physics. 

The value of the Cabibbo angle arises from an interplay of symmetry and symmetry breaking: is the value of the Cabibbo angle sitting in a natural minimum?

Unfortunately, Cabibbo and I found that natural minima are always trivial, i.e. correspond to Cabibbo angle $\theta_C=0,\pi$. 

The recent good news is that with the lepton flavor group (\ref{gylept}) it is possible to find~\cite{Alonso:2013nca} large mixing angles and correspondingly degenerate neutrinos\footnote{Majorana neutrinos may be degenerate in mass and still give rise to non trivial mixing, see~\cite{branco}.} unlike the hyerarchical situation and small angle found for quarks  (previous work with two generations was presented in~\cite{Alonso:2013mca}).

An independent approach leading to large neutrino mixing angles postulates a symmetry under discrete groups. We mention this possibility for completeness, referring the interested reader to the reviews in~\cite{altaferr,kingluhn}.

\vskip0.5cm

\subsection{\bf Natural minima of an invariant potential} We consider a potential $V(x)$ which is a function of certain fields $x$ that transform as a multiplet of a group ${\cal G}$. 

The potential must be a function of the independent invariants of the group,  $I_{i}(x)$, that we can construct out of $x$ fields. 
We expect to find the same number of invariants as the number of independent variables $x$. 

The key point is that the fields, i.e. the variables, span an entire space, while the manifold $M$ spanned by the invariants $I_{i}(x)$ has boundaries\footnote{in the case of rotations, $x$ is the coordinate and the invariant is $r^2$, the spatial coordinates go from $- \infty$ to $+\infty$ but the invariant $r^2$ cannot be negative.}.
 
\begin{figure}[!ht]
        \centering
       \includegraphics[scale=0.4]{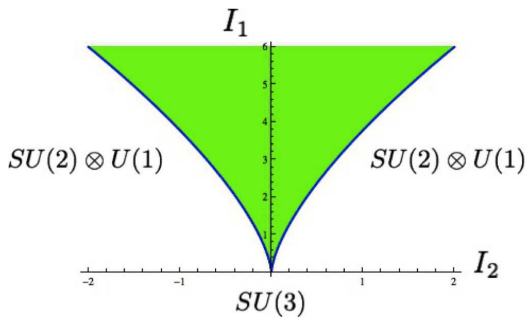}
  	   \caption{\label{InvariantManifold} Manifold spanned by the invariants of ${\cal G}= SU(3)$. Figure from Ref.~\cite{Alonso:2013nca}.}
\end{figure}
The manifold illustrated in~Fig. \ref{InvariantManifold} refers to ${\cal G}= SU(3)$, with $x$ belonging to the octet representation, namely an hermitian, $3 \times 3$, traceless matrix.

We may always take $x$ to be a diagonal matrix, which, due to the vanishing trace condition, has two independent eigenvalues. 
In correspondence, there are two invariants:
\begin{equation}
I_1={\rm Tr}(x^2), \;\;\; I_2={\rm Det}(x).
\end{equation}
and the boundary is represented by
 
\begin{equation}
I_1 \geq (54~I_2^2)^{1/3}~, \qquad-\infty < I_2 < +\infty~.
\label{boundary}
\end{equation}


It is possible to show that each boundary of $M$ corresponds to a subgroup of ${\cal G}$. For $SU(3)$ the points on the boundary lines are 
invariant under $SU(2) \otimes U(1)$, while the singular point, $x=0$, corresponds to invariance under the full $SU(3)$. 

The result found in~\cite{Cabibbo:1970rza} is that the {\it {\bf natural solutions for the minimum of the potential are found on the boundaries}}.

In the general case, the boundary of $M$ is made of ``surfaces'', joined by ``lines'' which converge on discrete ``points'' and the boundaries are identified by a very simple criterion. 

Consider the Jacobian matrix between the invariants and the fields:
\begin{equation}\label{Jacobian}
J=\frac{\partial(I_1,I_2,\cdots)}{\partial(x_1,x_2,\cdots)},
\end{equation}
One finds that on the boundary, the rank of $J$ has to be less than the dimension $N$ of the manifold $M$.
For instance, if rank$(J)= N-1$ we have a ``surface'', rank$(J)= N-2$ we have a ``line'', rank$(J)= N-3$ we have a ``point''.

We can explain this result in a simple way referring to Fig. \ref{InvariantManifold}. 

Consider a point on the boundary of $M$. A first order variation of $x$ induces a shift of this point. The shift cannot be orthogonal to the boundary because with a variation of $x$ of opposite sign the point would go outside the manifold $M$, which is impossible. Therefore, any first  order variation of $x$ must leave the starting point on the boundary. In particular, for $x=0$ any first order variation must leave the point unchanged, i.e. $x=0$ is always a stationary point. 

The extrema of $V(x)$ are to be found by solving the equations:
\begin{equation}
\frac{\partial V}{\partial x_j}=\sum_i \frac{\partial V}{\partial I_i}\frac{\partial I_i}{\partial x_j}=\sum_i \frac{\partial V}{\partial I_i} J_{ij}=0~.
\end{equation}

On the basis of the prevous consideration,  we may state that {\it {\bf the extrema of $V$ with respect to the points of a given boundary are extrema of $V(x)$}}~\cite{Cabibbo:1970rza}.

The latter extrema are more natural than the generic extrema in the interior of $\cal M$, since they require the vanishing of only $N-1$, or $N-2$, etc. derivatives of $V$ given that, on the boundary, $J$ has 1 or 2, etc. vanishing eigenvectors (i.e. the vectors orthogonal to the boundary). 

Thus, from Fig.~\ref{InvariantManifold} we learn that it is more natural to break $SU(3)$ along the direction of the hypercharge ($x$ with two equal eigenvalues, little group $SU(2)\otimes U(1)$) than along the direction of $T_3$, which corresponds to elements in the interior of $\cal M$~\cite{Michel:1970mua}.

In conclusion, we have a very nice criterion to find the natural extrema of $V(x)$: compute (\ref{Jacobian}) and find the surfaces of reduced rank.

In~\cite{Cabibbo:1970rza} we considered chiral $SU(3) \otimes SU(3)$, where $x$ stand for the quark masses and it was found that the 
natural extrema corresponded always to degenerate or hierarchical patterns, i.e
\begin{equation}
SU(3):~x=\left(\begin{array}{ccc}m & & \\ & m & \\ & & m\end{array}\right),
\end{equation}
or
\begin{equation}
SU(2)\otimes SU(2):~x=\left(\begin{array}{ccc}0 & & \\ & 0 & \\ & & m\end{array}\right).
\end{equation}
This means that the Cabibbo angle is always zero because, by an $SU(2)$ rotation, which is a symmetry of these minima, the symmetry breaking can always be aligned with the weak interactions.

\subsection{\bf The quark case with three families}
The quark case with the flavor group ${\cal G}_{quark}$, Eq.~(\ref{gflav}) is more complicated because we have many variables. 

The Lagrangian is
\begin{equation}
{\cal L}_Y={\bar Q}_L {\it Y}_U {\tilde H} U_R+{\bar Q}_L {\it Y}_D H D_R.
\end{equation}
The couplings are transformed as follows
\begin{eqnarray}
&& Y_U \to U_L Y_U U^U_R;~Y_D \to U_L Y_D U^D_R,\\
&& Y_U=diag=m_U;~Y_U=U_{CKM}\times diag=U m_D.\label{qdiag&ckm}
\end{eqnarray}
The independent parameters in quark Yukawa couplings are simply counted from eq.~(\ref{qdiag&ckm}): there are four parameters in the CKM matrix and six masses for a total of $10$ parameters. In correspondence, we may form $10$ independent invariants under the group ${\cal G}_{quark}$.

To classify these invariants, we define two matrices which transform in the same way under $SU(3)_q$ and are singlet under the
other transformations: 
\begin{equation}
\rho_U=Y_U Y_U^\dagger~, \qquad \rho_D=Y_D Y_D^\dagger~;  \qquad   \rho_{U,D} \to U_{q}  \rho_{U,D} U_{q}^\dagger~.
\end{equation}
There are six unmixed invariants, which we may take as:
\begin{equation}
I_{U^{1}}={\rm Tr}(Y_U Y_U^\dagger)~, \quad   I_{U^{2}}={\rm Tr}[(Y_U Y_U^\dagger)^2]~, \quad I_{U^{3}}={\rm Tr}[(Y_U Y_U^\dagger)^3]~,
\end{equation}
and the same for $Y_D Y_D^\dagger$. Next we define four mixed invariants:
\begin{equation}
\begin{array}{ll}
 I_{U,D}={\rm Tr}( Y_U Y_U^\dagger Y_D Y_D^\dagger)~, \quad 
& I_{U^2,D}={\rm Tr}[(Y_U Y_U^\dagger)^2 Y_D Y_D^\dagger)~,   \\
 I_{U,D^2}={\rm Tr}[Y_U Y_U^\dagger (Y_D Y_D^\dagger)^2]~, \quad 
& I_{(UD)^2}={\rm Tr}[( Y_U Y_U^\dagger Y_D Y_D^\dagger)^2]~. 
\end{array}\label{mixQ}
\end{equation}
 As anticipated, 10 independent invariants suffice to characterize in generality the physical degrees of freedom in
the Yukawa fields. We stress  in particular that the 4 invariants in Eq.~(\ref{mixQ}) contain enough information to reconstruct the 4 physical parameters of
the CKM matrix, including its CP-violating phase (up to discrete choices, see Ref.~\cite{Jenkins:2009dy}), despite none of them 
vanishes in the limit of exact CP invariance.

Now we proceed to classify the natural extrema.

As in the previous, chiral symmetry, case, unmixed invariants produce extrema corresponding to degenerate or hierarchical patterns ($m_u=m_c=0$, $m_t$=any value). 

Mixed invariants, that were not present in the previous analysis,  involve the CKM matrix $U_{CKM}$, e.g.
\begin{equation}
{\rm Tr}(Y_U Y_U^\dagger Y_D Y_D^\dagger)= \sum_{ij}U_{ij}U^\star_{ij}(m_U)_i(m_D)_j=\sum_{ij}P_{ij}(m_U)_i(m_D)_j ,
\end{equation}
and $P$ is what matematicians call a {\it bistochastic matrix}, i.e. a matrix where the sum of elements of any row equals the sum of elements of any column, with both sums equal to one. A theorem due to  Birkhoff and Von Neumann~\cite{vonneumann} states that the extrema of bistochastic matrices are permutation matrices. 

Therefore, the extrema of the mixed invariants are also permutation matrices. 
This means that there was a mistake in labelling quarks: relabelling the down quark 
coupled to each up quark we force the permutation matrix to be the unit matrix and we find again the hierarchical results we have quoted above~\cite{Cabibbo:1970rza}.

\subsection{
{\bf The lepton case with three families and see-saw}
}
In the lepton case with ${\cal G}_{lept}$ in eq. (\ref{gylept}), we recall\footnote{I am here following almost verbatim the discussion of Ref.~\cite{Alonso:2013nca}.}:
\begin{equation}\begin{split}
&{\cal L}_Y={\bar L}_L {\it Y}_E H E_R+\frac{1}{M}(\bar L_L {\it Y}_\nu {\tilde H}{\tilde H} {\it Y}_\nu^T L_L^c),  \\
& Y_E=y_E;~Y_\nu=U_L y_\nu  \omega  U_R;~y_E,y_\nu={\rm diagonal~matrices}.\label{formstandlept}
\end{split}\end{equation}
and neutrino masses are given by
\begin{equation}\label{masses}
M_\nu=\frac{v^2}{M}U_L(y_\nu \omega) U_R U_R^T( y_\nu \omega)U_L^T=U_{PMNS} \Omega~m_\nu~\Omega ~U_{PMNS}^T,
\end{equation}
$\Omega$ is the diagonal Majorana-phase matrix.


We need to construct $15$ independent invariants, Sect.~\ref{seesawnu}. We consider first the two combinations:
\begin{equation}
\rho_E=Y_E Y_E^\dagger~, \qquad  \rho_\nu= Y_\nu Y_\nu^\dagger~;  \qquad   \rho_{E,\nu} \to U_{\ell}  \rho_{E,\nu} U_{\ell}^\dagger~, 
\end{equation}
in which ${\cal O}(3)$ transformations disappear. We may construct unmixed and mixed invariants, as in the quark case, the mixed ones involving the matrix $U_L$, Eq.~(\ref{formstandlept}). We choose the unmixed ones as:
\begin{equation}
{\rm Unmixed, E:} \qquad 
I_{E^1}={\rm Tr}( Y_E Y_E^\dagger)~, \quad 
I_{E^2}={\rm Tr}[(Y_E Y_E^\dagger)^2]~, \quad 
I_{E^3}= {\rm Tr}[(Y_E Y_E^\dagger)^3]~,
\label{unmixE}
\end{equation}
and three similar ones  ($I_{\nu^{1-3}}$) using $\rho_\nu$,  while the four mixed invariants containing $\rho_E$ and $\rho_\nu$
are taken to be:
\begin{equation}
{\rm Mixed, ~type ~1}: \qquad 
\begin{array}{ll}
 I_{\nu,E}={\rm Tr}( Y_\nu Y_\nu^\dagger Y_E Y_E^\dagger )~, \quad  
 &  I_{\nu^2,E}={\rm Tr} [( Y_\nu Y_\nu^\dagger)^2 Y_E Y_E^\dagger ]~,  \\
 I_{\nu,E^2}={\rm Tr} [Y_\nu Y_\nu^\dagger (Y_E Y_E^\dagger)^2 ]  ~, \quad 
 & I_{(\nu E)^2}={\rm Tr}[( Y_\nu Y_\nu^\dagger Y_E Y_E^\dagger )^2] ~.
 \end{array}
\label{mixedenu}
\end{equation}

For neutrinos we may construct also a matrix which transforms under the orthogonal group only:
\begin{equation}
\sigma_\nu= Y_\nu^\dagger Y_\nu~;   \qquad   \sigma_\nu  \to { \cal O}  \sigma_\nu {\cal O}^T~.
\end{equation}
The symmetric and antisymmetric parts of $\sigma_\nu$ transform separately and can be used to construct two different invariants, such as 
${\rm Tr}[   Y_\nu^\dagger Y_\nu  (  Y_\nu^\dagger Y_\nu  \pm  Y_\nu^T  Y_\nu^* )]$. 
Here the first term in the product gives back the invariant  $I_{\nu^2}={\rm Tr} [( Y_\nu Y_\nu^\dagger)^2]$, but the second one gives rise to new contractions which involve the unitary, symmetric matrix 
\begin{equation}
W=U_R U_R^T~.
\label{blocco}
\end{equation}
We thus define the following three additional invariants:
\begin{equation}
{\rm Mixed, ~type ~2}: \qquad 
\begin{array}{ll}
J_{\sigma^1}={\rm Tr}(Y_\nu^\dagger Y_\nu Y_\nu^T  Y_\nu^* )~, \quad 
& J_{\sigma^2}={\rm Tr}[(Y_\nu^\dagger Y_\nu)^2 Y_\nu^T  Y_\nu^* ]~, \\ 
 J_{\sigma^3}={\rm Tr}[(Y_\nu^\dagger Y_\nu Y_\nu^T  Y_\nu^*)^2]~. & 
\end{array}
 \label{LRnu}
 \end{equation}
 Finally, we add two invariants which contain both $U_L$ and $W$:
 \begin{equation}
{\rm Mixed, ~type ~3}: \qquad 
\begin{array}{ll}
I_{LR}=\mbox{Tr}\left[{\it Y}_\nu {\it Y}_\nu^T {\it Y}_\nu^* {\it Y}_\nu^\dagger {\it Y}_E {\it Y}_E^\dagger \right]~, \\
I_{RL}=\mbox{Tr}\left[{\it Y}_\nu {\it Y}_\nu^T{\it Y}_E^* {\it Y}_E^T  {\it Y}_\nu^* {\it Y}_\nu^\dagger {\it Y}_E {\it Y}_E^\dagger \right]~.
\end{array}
\qquad\qquad 
\end{equation}

The discussion of the Jacobian leads to the following results, see Ref.~\cite{rodrigothesis} for details.
\begin{itemize}
 \item Unmixed invariants produce extrema corresponding to degenerate or hierarchical mass patterns.
  \item Mixed, type 1, invariants contain $|(U_L)_{ij}|^2$ and lead, like in the quark case, to the conclusion that $U_L$ is a permutation matrix 
 (up to an overall phase).
\item Mixed, type 2, invariants contain  $|W_{ij}|^2$ and indicate that $W=U_R U_R^T$ is also a permutation matrix  (up to an overall phase).
\item Once we impose that $U_L$ and $W$ are permutation matrices, the sensitivity of Mixed, type 3 invariants to $\omega$ vanishes. 
The latter remains therefore undetermined. 
 \end{itemize}
We may absorb the first permutation matrix in a relabeling of the neutrinos coupled to each charged lepton, 
but the second matrix may then lead to a non trivial result for the neutrino mass matrix, Eq.~(\ref{masses}). The reason for the difference is that, for quarks we could eliminate any complex matrix $U_D$ by a redefinition of $D_R$, but this is not possible for leptons, because we can redefine the $N_i$ only with a real orthogonal matrix.

We use the freedom in the neutrino labeling to set $U_L=1$ in the basis where charged leptons are ordered according to: 
\begin{eqnarray}
{\it Y}_E=\mbox{diag}\,(y_e, y_\mu,y_\tau)~.
\label{chlept}
\end{eqnarray}
There are four possible symmetric permutation matrices that can be associated with $W=U_R U_R^T$, one of them being the unit matrix. The other three imply non trivial mixing in one of the three possible neutrino pairs, e.g.
\begin{eqnarray}
W=U_R U_R^T=-\left( \begin{array}{ccc} 1&0&0\\0&0&1\\0&1&0 \end{array}\right)
\end{eqnarray}
We introduced the minus sign for $W$ to have a positive determinant, consistently with the  condition ${\rm Det}(U_R)=1$. 

Using this expression in Eq.(\ref{masses}) leads to 
\begin{eqnarray}
m_\nu=\frac{v^2}{M}~y_\nu \omega  W  \omega  y_\nu=\frac{v^2}{M}~\left(\begin{array}{ccc}-y_1^2 e^{2i\alpha} & 0 & 0\\ 0 & 0 & -
y_2 y_3 e^{-i\alpha}  \\0 &- y_2 y_3 e^{-i\alpha}  & 0 \end{array}\right)~,
\label{solution}
\end{eqnarray}
where $y_\nu=$~diag($y_1,y_2,y_3$) and $\omega$= diag($e^{i\alpha} , e^{i\beta} , e^{-i(\alpha+\beta}$).  
The absence of mixing between the first eigenvector of $m_\nu$ and those associated to the 2-3 sector implies that the 
phase $\alpha$ is unphysical and can be set to zero by an appropriate phase redefinition of the neutrino fields.
From the second equality in Eq.~(\ref{masses}) we then find:
\begin{eqnarray}
&& {\hat m}_\nu=\frac{v^2}{M}~{\rm diag}(y_1^2,y_2y_3,y_2y_3)~,\notag \\
&&U_{\rm PMNS}^{(0)}=\left(\begin{array}{ccc}1 & 0 & 0\\ 0 & 1/\sqrt{2} & 1/\sqrt{2}\\ 0 & -1/\sqrt{2} & 1/\sqrt{2} \end{array}\right)~, 
\qquad \Omega= {\rm diag}(-i,-i,1)~.
\label{pmns0}
\end{eqnarray}
The non-trivial Majorana phase difference in the 2-3 sector
 is needed to bring all masses in positive form. There are one maximal mixing angle and one maximal  Majorana phase, which stem from the ${\cal O}(2)$ substructure in Eq.~(\ref{pmns0}), as found in Ref.~\cite{Alonso:2012fy}.

With three families we can go closer to the physical reality if we assume complete degeneracy for $y_\nu$. In this case, after the $2-3$ rotation we are left with degenerate $1$ and $2$ neutrinos and, a priori, a new rotation will be needed to align the neutrino basis with the basis in which the charged lepton mass takes the diagonal form in Eq.~(\ref{chlept}). We may expect, in this case, the PMNS matrix to have an additional rotation in the $1-2$ plane:
\begin{equation}
U_{\rm PMNS}=U_{\rm PMNS}^{(0)} ~U(\theta_{12})~.
\end{equation}
We shall see that small perturbations around the solution in Eq.~(\ref{solution}) allow to determine this angle, that remains non-zero 
in the limit of vanishing perturbations.

\subsection*{ {\bf Group theoretical considerations} }
One may ask what is the little group corresponding to the extremal solution, Eq.~(\ref{solution}). While $Y_\nu$ transforms under $SU(3)_{\ell}\otimes {\cal O}(3)$, orthogonal transformations drop out of ${\it Y}_\nu{\it Y}_\nu^T$. In some sense we have to find the appropriate square root of $m_\nu$.  
By explicit calculation, one sees that the answer is given by\footnote{${\it Y}_\nu$ is uniquely determined up to an inessential right multiplication by an orthogonal matrix.}:
\begin{eqnarray}
&&Y_\nu=
\left(\begin{array}{ccc}i y_1 & 0 & 0 \\ 0 &i \frac{y_2}{\sqrt{2}} &  \frac{y_2}{\sqrt{2}} \\ 0 & i \frac{ y_3}{\sqrt{2}} &-\frac{y_3}{\sqrt{2}} \end{array}\right)~.
\label{squareroot}
\end{eqnarray}
${\it Y_\nu}$ transforms under $SU(3)_{\ell}\otimes {\cal O}(3)$ according to the $(\bar 3,3_V)$ representation, where the suffix V denotes the vector representaton of $\mathcal O (3)$, realized, in triplet space, by the Gell-Mann imaginary matrices $\lambda_{2,5,7}$. One verifies that:
\begin{equation}
\lambda_3^\prime Y_\nu-Y_\nu \lambda_7=0;~\lambda_3^\prime={\rm diag}(0,1,-1)~,
\end{equation}
i.e.~for this solution, $SU(3)_{\ell}\otimes {\cal O}(3)$ is reduced to the $U(1)_{\rm diag}$ subgroup of transformations of the form:
\begin{equation}
U(1)_{\rm diag}:~{\rm exp} \left(i\epsilon \lambda_3^\prime\right)\otimes {\rm exp}\left(i\epsilon\lambda_7\right)~.
\end{equation}
 This $U(1)_{\rm diag}$ is the little group of the boundary to which the solution in Eq.~(\ref{squareroot}) belongs. 
When combined with a hierarchical solution for the charged-lepton Yukawa of the type $Y_E\propto (0,0,1)$,
this corresponds to the little group $SU(2)_E \otimes U(1)_{\rm diag}$,
a subgroup of $SU(3)_\ell \otimes SU(3)_E \otimes {\mathcal O(3)}$.

In the limit  $y_1= y_2 =y_3$, ${\it Y}_\nu$ becomes proportional to a unitary matrix:
\begin{eqnarray}
&&Y_\nu \to y~
\left(\begin{array}{ccc}i & 0 & 0 \\ 0 & i \frac{1}{\sqrt{2}} & \frac{1}{\sqrt{2}} \\ 0 & i \frac{1}{\sqrt{2}} &- \frac{1}{\sqrt{2}} \end{array}\right)
=y V~,
\qquad V V^\dagger =1~,
\label{eq:40}
\end{eqnarray}
and the $U(1)$ invariance is augmented to a full ${\cal O}(3)_{\rm diag}$, a maximal subgroup of $SU(3)_{\ell}\otimes {\cal O}(3)$:
\begin{equation}
Y_\nu\to (V{\cal O}V^\dagger) Y_\nu {\cal O}^T= Y_\nu~,
\end{equation} 
where ${\cal O}$ is an orthogonal matrix  generated by $\lambda_{2,5,7}$. The ${\cal O}(3)_{\rm diag}$ would remain 
unbroken only in the case of degenerate charged lepton masses. Combining $Y_\nu$ in  Eq.~(\ref{eq:40})
 with $Y_E\propto (0,0,1)$,
we recover the little group $SU(2)_E\otimes U(1)_{\rm diag}$. 

Summarizing:
\begin{itemize}
\item{} $Y_E\propto (0,0,1): \quad SU(3)_{E} \otimes SU(3)_\ell \to SU(2)_E\otimes SU(2)_\ell \otimes U(1) \quad$ (maximal~subgroup) 
\item{} $Y_\nu$ in  (\ref{squareroot})~:  $\qquad\quad  {\hat m}_\nu={\rm diag}(m_1,m,m)~,
\qquad~SU(3)_{\ell}\otimes {\cal O}(3)\to U(1)_{\rm diag}$
\item{} $Y_\nu$ in (\ref{eq:40})~: $\quad\quad {\hat m}_\nu=m\times 1~, \qquad  ~ SU(3)_{\ell}\otimes {\cal O}(3) \to {\cal O}_{\rm diag}(3)$
\quad  (maximal subgroup)
\item{} $Y_E\propto (0,0,1)$ \& $Y_\nu$ in (\ref{squareroot}) or (\ref{eq:40})~:
  $\quad\ SU(3)_{E} \otimes SU(3)_{\ell}\otimes {\cal O}(3) \to  SU(2)_E\otimes U(1)_{\rm diag}$
\end{itemize}
Both breaking patterns of $Y_\nu$ feature: i) at least two degenerate neutrinos; ii)~$\theta_{23}=\frac{\pi}{4}$ and $\theta_{13}=0$;
iii)  one real and one imaginary Majorana phases. In addition, the degenerate pattern in Eq.~(\ref{eq:40})  implies 
three degenerate neutrinos and a second large (not calculable) mixing angle.



\subsection{\bf Perturbations}
We may consider what happens when we introduce small perturbations around this particular solution, i.e.
\begin{equation}
M_\nu = \frac{v^2 y}{M}\left(\begin{array}{ccc}1+\delta & \epsilon+\eta & \epsilon-\eta\\\epsilon+\eta & \delta & 1\\\epsilon-\eta & 1& 
\delta \end{array}\right).
\end{equation}
To first order in perturbations we find
\begin{eqnarray}
&&m_\nu = m\left(\begin{array}{ccc}1+\delta+\sqrt{2}\epsilon & 0 &0 \\ 0 & 1+\delta-\sqrt{2}\epsilon & 0 \\0 & 0 & -1+\delta \end{array}\right), \label{Diag_neutrino_mass}\\ 
\nonumber \\
&&U_{PMNS}=\left(\begin{array}{ccc}
1/\sqrt{2} & -1/\sqrt{2} & \eta/\sqrt{2}\\ 1/2(1+\eta/\sqrt{2}) & 1/2(1-\eta/\sqrt{2}) & -1/\sqrt{2} \\ 1/2(1-\eta/\sqrt{2}) & 1/2(1+\eta/\sqrt{2}) & 1/\sqrt{2} \end{array}\right). \label{PMNS}
\end{eqnarray}
The PMNS matrix has the smallest entries that correspond to $\theta_{13}$, which in fact is much smaller than the others.
If we estimate $\sin\theta_{13}$ or, equivalently, the deviation of $\theta_{12}$ from $\pi/4$ and assume that they are of the same order 
of the perturbations, we have
\begin{equation}
\frac{|\Delta m_{atm}^2|}{2m_0^2} \approx |sin \theta_{13}| \approx |\theta_{12}-\frac{\pi}{4}| \approx 0.1 ~\to~ m_0 \approx 0.1 {\rm eV}.
\label{estimate}\end{equation}

In this approximation we find that neutrinos masses \eqref{Diag_neutrino_mass} are essentially degenerate.  

Extrapolating from quarks we would expect small mixing angles and a hierarchical pattern, with $|\Delta m^2|\sim m^2$, $m$ being the mass of the heaviest quarks . 

\begin{figure}[!ht]
        \centering
       \includegraphics[scale=0.3]{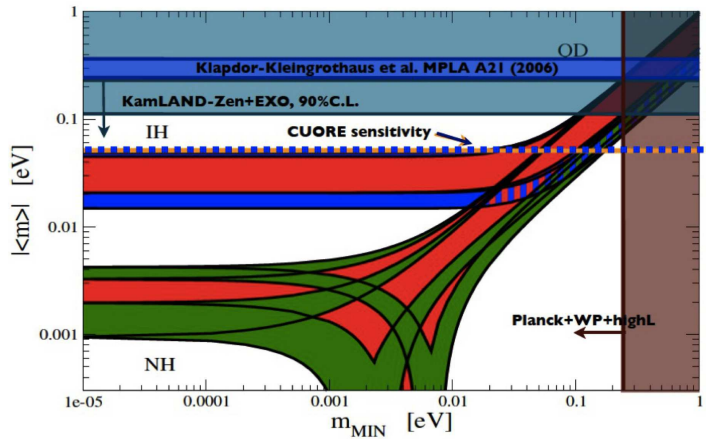}
  	   \caption{\label{PredictionsDbdecay} Neutrino masses coming from double beta decay without neutrinos. Courtesy of S. Pascoli~\cite{pascoli}.  }
\end{figure}

At variance with the quark case, extremizing a potentia invariant under the neutrino flavor symmetry one finds a natural solution with large mixing angles and a degenerate mass pattern. The mass estimated in (\ref{estimate}) would lead to a rate for neutrinoless  double beta decay not too far from the present limits, as shown in Fig.~\ref{PredictionsDbdecay}. 

Three almost degenerate neutrinos with $m_0\sim 0.1$~eV would be compatible with the recent value of the sum of neutrino  masses reported by the Planck Collaboration~\cite{Ade:2013lmv}, on the basis of cosmological data:

\begin{equation}
\sum m_\nu = 0.22 \pm 0.09~ {\rm eV}.
\end{equation}

\vskip0.5cm

\section{{ Outlook}}
\label{outlook}

Large as it has been the progress of the last decades in neutrino physics, we may still list a good number of open issues.

Given that oscillations determine only the magnitude of  the mass-squared differences, a first relevant problem is the neutrino mass hierarchy, what is the masss ordering of $\nu_{1,2,3}$ and whether neutrino masses are largely spaced, as is the case of quarks, or are almost degenerate, with small mass differences. The different textures of CKM and PMNS matrices makes it suspicious to assume a similarity in the mass spectrum. 

If neutrinos are Majorana particles, detection of neutrinoless  double beta decay could give the crucial information. 
Cosmological observations seem also to be close to a determination of the absolute value of neutrino masses, see e.g.~\cite{Ade:2013lmv},  if they are indeed much larger than the corresponding mass differences.

The observation of the CP violating phase is also a problem of paramount importance, and encouraging good news is the relatively large value of $\sin \theta_{13}$,  which always multiplies the CP violating phase in the PMNS matrix (\ref{pmatrix}).

On the theory side, the challenge is to find an explanation of the difference between flavor violation in quarks and leptons and, even more, to find a theoretical path to the calculation of the mixing angle, the Cabibbo's dream of the sixties.

We have explored the idea that Yukawa couplings satisfy a minimum principle with a potential symmetric under the flavor group of the Standard Theory.
The existence of three fermion generations and  heavy Majorana neutrinos leads to two interesting solutions: (i) hierarchical mass pattern and unity CKM matrix for quarks; (ii)  hierarchical masses for charged leptons, almost degenerate Majorana neutrinos with one, potentially two, large mixing angles.

Both solutions are close to the real situation. The prediction that large mixing angles are related to Majorana degenerate neutrinos may 
be amenable to experimental test in a not too distant future. 

Formulating the idea within a renormalizable theory at relatively low energy requires special care to be consistent with experiments~\cite{Grinstein:2010ve}, or maybe the new fields live at very high energies, as supposed originally in~\cite{Froggatt:1998tj}. 

Future will tell if those presented here are fruitful ideas or simply a dream with open eyes.

\vskip0.5cm

\section*{{ Acknowledgements}}

I am indebted to E. Fiorini for the many illuminating talks he has given on double beta decay, which provided me with useful concepts and  innumerable figures and diagrams. Conversations with F. Feruglio, B. Gavela, A. Melchiorri, S. Pascoli and A. Polosa are gratefully acknowledged. I am grateful to M.  Paolella and E. Battista for efficiently providing  the first draft of the article out of rather complicated slides. Finally, I would like to acknowledge the hospitality of Istituto de Fisica Teorica, Universidad Autonoma de Madrid, where most of this article was written.
\section*{APPENDIX I}

\paragraph{\bf A little hystory first} After the discovery of parity and strangeness violation, there were suggestions that terms with these properties could appear in the quadratic part of the lagrangian, due to higher order interactions. 
For mass terms, one could consider, for example, the form:
\begin{equation}
{\cal L}={\bar \psi}(A+iB\gamma_5)\psi
\label{l&pviolating}
\end{equation}
with 
\begin{equation}
\psi=\left(\begin{array}{c} e \\ \mu \end{array}\right)
\end{equation}
and $A$ and $B$ non-diagonal, hermitian matrices, to preserve the hermiticity of ${\cal L}$. ${\cal L}$ appears to violate parity and lepton number conservation but what is the meaning of this violation?

The answer was given by Cabibbo and Gatto~\cite{cabgat} and by Kabir, Feinberg and Weinberg~\cite{kabfeinwein} who showed that fields could be redefined so as to transform away this term into a canonical, diagonal mass term of the form:
\begin{equation}
{\cal L}_{mass}={\bar \psi}M\psi
\label{massterm}
\end{equation}
with $M\geq 0$. 

Of course, the same field redefinition has to be carried over in the other terms of the lagrangian and this would transfer the implied parity and lepton number violation to the interaction. Note that this is exactly what is done following eq.~(\ref{ckmdef}) for what concerns flavor violation.

To connect to eq.~(\ref{theorem}), we introduce left-and right-handed fields, with $\psi=\psi_L +\psi_R$. Eq.~(\ref{l&pviolating}) now reads:
\begin{equation}
{\cal L}={\bar \psi_L}(A+iB)\psi_R+{\bar \psi_R}(A-iB)\psi_L={\bar \psi_L}{\cal M}\psi_R~+~{\rm h.c.}
\label{chiral}
\end{equation}
in terms of a generically complex, non-diagonal matrix ${\cal M}$. By performing the field redefinition (which leaves unchanged the canonical anticommutation relations):
\begin{equation}
\psi_L\to U\psi_L;~\psi_R\to V \psi_R
\label{canontransfg}
\end{equation}
${\cal M}$ transforms as:
\begin{equation}
{\cal M}\to U{\cal M}V^\dagger=M
\label{canontransfg}
\end{equation}

The theorem stated in eq.~(\ref{theorem}) tells that with an appropriate choice of $U$ and $V$, $M$ is diagonal, real and positive and (\ref{l&pviolating}) is reduced to (\ref{massterm}).

Now we can prove the theorem stated in eq.~(\ref{theorem}), following Ref.~\cite{cabgat}.

\paragraph{\bf Dim} We restrict to the case where ${\cal M}$ is non singular. The case of one one more vanishing eigenvalues is treated by continuity from the non singular case.

We start by defining the two matrices:
\begin{equation}
{\cal M M^\dagger}=H_a;~{\cal M^\dagger M}=H_b
\end{equation}

$H_{a,b}$ are both hermitian, positive definite and we prove that they have the same eigenvalues. Indeed they satisfy the same secular equation:
\begin{eqnarray}
&&0=det\left({\cal M M^\dagger}-\lambda\right)=det\left[{\cal M}\left({\cal M^\dagger}-\lambda{\cal M}^{-1}\right)\right]=det\left[\left({\cal M^\dagger}-\lambda{\cal M}^{-1}\right){\cal M}\right]=\nonumber \\
&&=det\left({\cal M^\dagger M}-\lambda\right)\nonumber
\end{eqnarray}
Therefore, there exist two unitary matrices, $U$ and $V$ such that:
\begin{equation}
U H_a U^\dagger=V H_b V^\dagger=\sigma
\end{equation}
with $\sigma$ diagonal and positive.

Using $U$ and $V$, we construct the matrices:
\begin{eqnarray}
&&h=U{\cal M}V^\dagger,~h^\dagger=V{\cal M^\dagger}U^\dagger;\nonumber
\end{eqnarray}
and note that
\begin{eqnarray} 
&& h h^\dagger=U{\cal M M^\dagger}U^\dagger=\sigma \nonumber \\
&&h^\dagger h=V{\cal M^\dagger M}V^\dagger=\sigma\nonumber
\end{eqnarray}
evidently, $h$ and $h^\dagger$ commute and we may treat them as numbers.

In particular, one sees immediately that:
\begin{itemize}
\item $h(h^\dagger)^{-1}=(h^\dagger)^{-1}h=\frac{h}{h^\dagger}=Z$ is a unitary matrix;
\item $h h^\dagger$ is hermitian positive
\item $h^2=h h^\dagger(h^\dagger)^{-1}h=h h^\dagger Z$
\item taking the square root, one has $h=H'Z'$, with $H'$ hermitian positive and $Z'$ unitary
\item finally, from $h=U{\cal M}V^\dagger$, we obtain: ${\cal M}=U^\dagger H' Z' V= U^\dagger H' U (U^\dagger Z' V)=H W$
\end{itemize}
with $H$ hermitian and positive and $W$ unitary.

We leave to the reader the derivation of the corollary, eq.~(\ref{corollary}).


\end{document}